\documentclass{article}
\usepackage{amsmath,amssymb}
\usepackage{epsfig,multirow}
\usepackage{xcolor}
\usepackage{bigstrut}
\usepackage{array}
\usepackage{hyperref}

\usepackage[a4paper,top=3cm,bottom=2cm,left=3cm,right=3cm,marginparwidth=1.75cm]{geometry}
\title{Case for Centaurus~A as the main source of ultrahigh-energy cosmic rays}

\author{Silvia Mollerach and Esteban Roulet\\
Centro At\'omico Bariloche, Comisi\'on Nacional de Energ\'\i a At\'omica\\
Consejo Nacional de Investigaciones Cient\'\i ficas y T\'ecnicas (CONICET)\\
Av. Bustillo 9500, R8402AGP, Bariloche, Argentina}
\date{}

\begin{document}
\maketitle
\begin{abstract}
    We discuss the possibility that a dominant fraction of the cosmic rays above the ankle (i.e. above 5~EeV) is   due to a single nearby source, considering in particular the radio galaxy Centaurus~A. We focus on the  properties of the source spectrum and composition  required to reproduce the observations, showing that the nuclei are strongly suppressed  for $E>10\,Z$~EeV, either by a rigidity dependent source cutoff or by the photodisintegration interactions with the CMB at the giant dipole resonance. The very mild attenuation effects at lower energies imply that the secondary nuclei from this source, produced in photodisintegration processes during propagation, only provide a small contribution.  Given the moderate anisotropies observed, the  deflections in extragalactic and Galactic magnetic fields should play a crucial role in determining the cosmic ray arrival direction distribution. The diffusion in extragalactic fields as well as the finite source lifetime also significantly affect the shape of the observed spectrum. The cosmic ray flux at tens of EeV is dominated by the CNO component, and we show that it is actually better reproduced by a mixture of C and O nuclei rather than by the usual assumption of  a N component effectively describing this mass group. The Si and Fe group components become dominant  above 70~EeV, in the energy range  in which a strong spectral suppression is present. If the localised flux excess appearing above 40~EeV around the Centaurus~A direction is attributed to the CNO component, the He nuclei from the source in the energy range from 10 to 20~EeV could lead to a similar anisotropy unless its contribution is suppressed. The cosmic ray flux at a few EeV should  mostly result from a more isotropic light component associated to a population of extragalactic sources. The inclusion of the subdominant contribution of heavy nuclei from the Galactic component helps to reproduce the  observations around 1~EeV.

\end{abstract}
\section{Introduction}

The spectrum of the ultra-high energy cosmic rays (UHECRs)  has been measured with great detail in recent years \cite{ab21,ve17}. Several features are present in it, such as the hardening at the energy of the ankle (at $\sim 5$~EeV), a subsequent steepening at the instep (at $\sim 13$~EeV) and a very strong suppression taking place beyond 50~EeV. On the other hand, the composition determined from the distribution of the depth of maximum development of the air showers, $X_{\rm max}$, suggests that at energies of a few EeV the cosmic rays (CRs) are relatively light, and  they become progressively heavier for increasing energies \cite{yu19}.
Moreover, beyond the ankle the distribution of their masses should be quite narrow at any given energy, given the small values observed of the dispersion of $X_{\rm max}$. 

Considering different astrophysical source scenarios and fitting the observations mentioned above, all these measurements at Earth have been used to infer the properties of the spectra of the CRs injected at the sources themselves \cite{al14,aa17b, ab23}. 
Two different source populations are actually required in order to account for the UHECR observations above $\sim 1$~EeV and across the ankle, one dominating the fluxes up to a few EeV and another one dominating at higher energies. Considering  that the sources are extragalactic and densely distributed in the Universe up to high redshifts and that each population consists of a mixture of different representative elements (such as H,  He, N, Si and Fe) emitted with power-law spectra having a rigidity dependent cutoff   (as expected from stochastic acceleration by electromagnetic processes), a fit to the measurements allowed one to infer the parameters that characterize the emitted fluxes \cite{ab23}. In particular, the low-energy population  (LE) dominating the fluxes below a few EeV
should consist mostly of light and intermediate mass nuclei (H, He, and N) having steep spectra proportional to $E^{-\gamma_{\rm L}}$ below the cutoff rigidity, with $\gamma_{\rm L}\simeq 3$ to 3.7, and the cutoff being only loosely determined given the steep nature of this spectrum. The high-energy population (HE)  dominating at higher energies  should instead be heavier, with an important contribution from N nuclei at few tens of EeV and  heavier nuclei dominating at the highest energies. The source spectrum of this HE component should be quite hard, proportional to $E^{-\gamma_{\rm H}}$ below the corresponding cutoff rigidity, with $\gamma_{\rm H}<1$ and in some scenarios its value even becoming negative. An important fact is that the cutoff rigidity of this component turns out to be relatively small, so that cosmic rays of charge $Z$ from the HE component have in general energies smaller than 10$Z$\,EeV. The interactions of these heavy CRs with the radiation backgrounds will shape the spectrum at Earth and will also produce sizeable amounts of H and He nuclei of  secondary origin, even if those elements may be emitted in low proportions at the sources themselves.

 In this context, the falloff of the He component can explain the instep feature observed, while the strong suppression above 50~EeV is due to the falloff of the N component. These suppressions are  due to a combination of the effects  of the rigidity cutoff at the  sources and of interactions  with background radiation during propagation. In the scenarios described above, the hardness of the spectra at Earth of the different elements of the HE population  could be associated   with a very hard spectrum emitted at the sources, either due to peculiar properties of the acceleration mechanism or as the result of interactions taking place around the sources \cite{gl15,un15}. Alternatively, this could be due to a magnetic horizon effect taking place as the CRs from a discrete distribution of sources propagate across strong intergalactic magnetic fields, in which case the diffusing low-rigidity CRs may not have enough time  to reach the Earth even from the closest sources  \cite{le05,be07,mo13,mo20,go23}. Still another possibility would be that the low rigidity CRs get magnetically confined in the galaxy clusters hosting the sources, effectively hardening the outcoming spectra \cite{ha16b}.

In this work we want to discuss in detail the implications of the spectrum and composition measurements for a different type of scenario \cite{mo19b, mo22}, in which the HE component arises mostly from  the contribution of a nearby extragalactic source at a distance of several Mpc, and focusing in particular in the case of the Centaurus~A radio galaxy (Cen~A, which is the radio counterpart of NGC~5128 \cite{is98}). This is the closest active galaxy and lies near the center of the localized excess of CRs hinted  on angular scales of about $30^\circ$ radius at energies larger than about 40~EeV \cite{aa15, go23b}, what makes it a particularly attractive UHECR source candidate. Also a combined fit of the spectrum, composition and arrival direction's information from the Auger Observatory  supports the presence of a flux excess at the highest energies associated to the direction of Cen~A \cite{ab24}.

While a cosmological distribution of sources emitting heavy nuclei up to the highest energies usually leads to a very large amount of light secondary particles (mostly H and He) at energies of a few EeV, 
a characteristic of the nearby source scenario  considered here is that the amount of secondary nuclei from this source resulting from photodisintegration processes will turn out to be relatively small. 
One may note that  the energies at which the different components appear to be suppressed, such as the He near the instep feature or the CNO elements near the suppression energy, are not very different from the energies at which these elements should be strongly suppressed due to the interactions with the  cosmic microwave background (CMB), so that it is tempting to try to associate the suppressions of the individual elements to interaction effects rather than to a fine-tuned coincidence of the values of the source cutoffs with the energies at which the suppression by interactions should appear. However,  given the steepness of the suppression present above 50~EeV and the lack of large amounts of  secondary hydrogen between the ankle and the instep, in the scenarios with continuous source distributions it turns out that the source cutoff cannot be much larger than the energies at which the suppression from the interactions is expected to appear, and it is indeed found to be just slightly smaller than it. In this context, one could also point out that if many sources were to contribute to the HE component,  the assumption that they all have a similar cutoff rigidity is not very natural \cite{eh23}, and hence the observation of a very narrow distribution of masses at each energy above the ankle is somewhat surprising if it does not arise from interaction effects which are universal or if there is just one single dominant source.

One important difference of the nearby source scenario is that the only relevant interaction of the nuclei as they propagate from the HE source through intergalactic space is eventually the one due to the disintegrations with the photons of the CMB. The interactions with the extragalactic background light (EBL) have associated   attenuation lengths which are much larger than the distance to the source, and at lower energies where the diffusion eventually happens they will  anyhow be smaller than the maximum distance travelled since the beginning of the source activity that we will consider. In the case of H nuclei, the pair production processes off CMB photons have also much larger attenuation lengths, and the photopion production off CMB photons, which eventually opens up at a threshold larger than about 70~EeV  (which will however turn out not to be achieved), could have a non-negligible impact only for sources farther than few tens of Mpc. For the case of sources distributed up to cosmological distances,  the photodisintegration interactions of nuclei  with the EBL are responsible for a significant steepening of the spectrum of the CRs reaching the Earth, which begins already one decade in energy below the energy at which  interactions with CMB photons become dominant. This is  largely responsible for the fact that  in those scenarios very hard emission spectra are inferred for the HE component in order to compensate for the steepening due to the interactions with the EBL. Note that those hard spectra are  in tension with the usual expectations from diffusive shock acceleration, that would require that the spectral index at the sources  be close to $\gamma= 2$, and hence the consideration of scenarios in which this may be avoided, such as the one with a powerful nearby source,  is of particular interest also in this respect.

This nearby source scenario can in principle  explain the presence of the localized excess around the direction of Cen~A observed above 40~EeV, and which  still appears to be present above 60~EeV. However, the presence of sizeable extragalactic magnetic fields \cite{va11,fe12} in the local neighbourhood containing  the source and the observer, as well as the magnetic lensing effects of the Galactic magnetic field component, should be important in order to explain the more isotropic distribution of arrival directions observed, which  at lower energies is essentially  characterized by an approximately dipolar distribution with a moderate amplitude (of about 7\% at 10~EeV) \cite{aa17,aa18}, and pointing in a direction somewhat displaced with respect to that of Cen~A. The magnetic deflections are indeed expected to play a very relevant role given the  low  rigidities that are inferred for the increasingly heavier UHECR  ($E/Z<10$~EeV). In particular, the deflections in the extragalactic magnetic fields (EGMF), with required strengths of several tens of nG in the Local Supercluster region in which magnetic fields are indeed expected to be enhanced, should contribute to the angular dispersion of the CR trajectories at the highest rigidities achieved and   lead to spatial diffusion   at lower rigidities, smoothing out significantly the anisotropies and also giving rise to a modulation of the CR spectrum. The Galactic  magnetic field  also plays a relevant role in the isotropization and reorientation of the flux reaching Earth.\footnote{A different explanation for the origin of the observed dipolar anisotropy is that it be due to the nonuniform distribution of the galaxies in a scenario in which many extragalactic sources contribute \cite{ha14,aa17,di21,al22}. } 
We will here describe the transition from the rectilinear to the diffusive regime  following refs.~\cite{ha14,ha16}.  Note also that the effects of interactions during propagation from the nearby source are only relevant at the highest rigidities, and will hence not differ significantly from those associated to straight trajectories, so that we will compute the attenuations in this approximation. 

Let us mention that scenarios with nearby UHECR sources dominating the flux above the ankle energy have been considered since long ago, see e.g. \cite{wd79,gi80,be90,bl99,fa00} where sources such as M87 or Cen~A were discussed. To avoid the presence of strong anisotropies a diffusive propagation in  extragalactic space had to be assumed, and the light proton composition considered by then required the presence of extragalactic magnetic fields typically larger than 100~nG in the Local Supercluster region.  Scenarios in which several nearby radio galaxies contribute to the observed fluxes above the ankle energy, considering also a mixed heavier composition,  were more recently analysed for instance in \cite{ma18,ei22}.

One should also keep in mind that a time dependent source emissivity will have an impact on the observed spectral features \cite{ha16,ei23}. This could happen if the source was  transient, as would for instance  be the case  if the source  was affected by outbursts related to episodes of enhanced accretion into the central supermassive black hole of the source galaxy or if it started its activity at some given time in the past after a special episode such as a merging of galaxies, as actually happened in the case of Cen~A. This will in particular lead to a  suppression of the spectrum at low rigidities, when the time of diffusive propagation from the source becomes larger than its lifetime. 

We will perform our study in progressive steps, starting  with the case of no magnetic deflections and considering 
 the source emission to be steady, what will allow us to understand several features of these scenarios which will also be relevant in the more realistic case. We will then introduce the effects of magnetic fields and of a finite source lifetime, study their impact on the spectra as well as on the distribution of arrival directions, to arrive to scenarios in which the different observations could be accounted for.

In order to accurately fit the features of the spectrum near the suppression energy, actually  some hints will be found in favor of the presence of  emitted C and O nuclei from the nearby source,  rather than the usual assumption of a  N component effectively describing the CNO group. Also the presence of a heavy Galactic component will help to better reproduce the composition measurements at energies below a few EeV.  Another important consequence that we will discuss is that if the CNO component of the nearby source is producing the localized excess observed above 40~EeV, then the He component from that source, be it of primary or of secondary   origin, should be subdominant in the 10 to 20~EeV range, since otherwise at these energies it could lead to a large anisotropy  in the same region of the sky.

\section{Cosmic-ray observables and general characteristics of the different source models}

When inferring the CR masses from the $X_{\rm max}$ measurements, it has to be noted that the actual mass composition derived depends on the hadronic interaction model used to interpret the air-shower development observed \cite{yu19}. In particular,  on the basis of the QGSJet model one infers a lighter composition than with the Sibyll model, since the first predicts that the air showers should be less penetrating.  The EPOS-LHC model leads in turn to  results which are intermediate between the previous two. On the other hand, the $X_{\rm max}$ fluctuations expected from each different CR element are less dependent on the hadronic model, and they become smaller as the masses increase. Hence, when small values of $\sigma(X_{\rm max})$ are observed, as is the case above the ankle energy, this  suggests that an almost pure heavy composition is present, since both a lighter composition or a mixture of different masses would tend to increase the fluctuations. This implies that an inconsistency can appear if the average $\langle X_{\rm max}\rangle$ value requires a lighter composition than the one expected from the fluctuations,  as  happens above the ankle when  the QGSJet model is considered. This is reflected  in the fact that when translating the $X_{\rm max}$ measurements in terms of the average $\langle {\rm ln}A\rangle$ and variance    Var$({\rm ln}A)$, with $A$ the  mass number of the CRs reaching Earth,  unphysical negative values for  Var$({\rm ln}A)$ are obtained above the ankle energy when considering the QGSJet model. We will hence not discuss this model and just show results relying on the composition based on the Sibyll~2.3c model that were obtained in \cite{yu19}, given also that the latest preliminary version of EPOS   \cite{pi23} gives predictions which are closer to the ones of the Sibyll model.

\begin{figure}[b]
    \centering
    \includegraphics[width=0.3\textwidth]{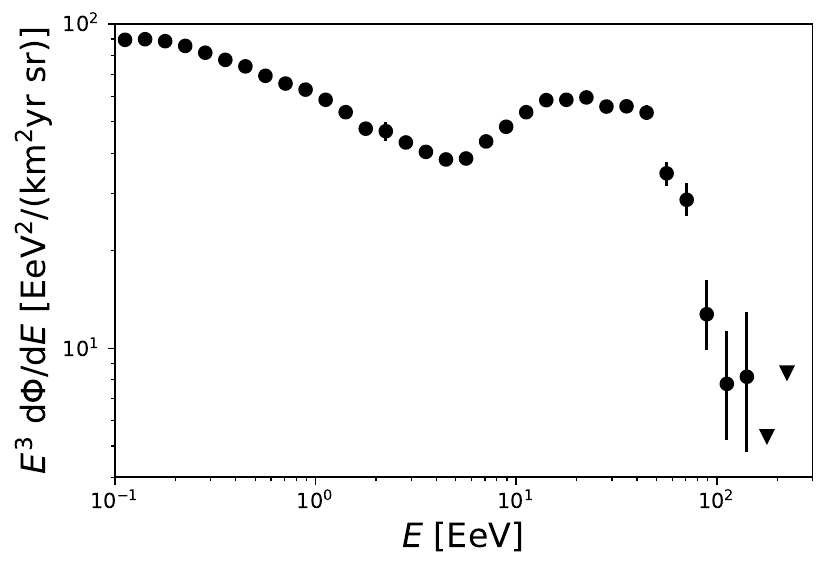}\includegraphics[width=0.3\textwidth]{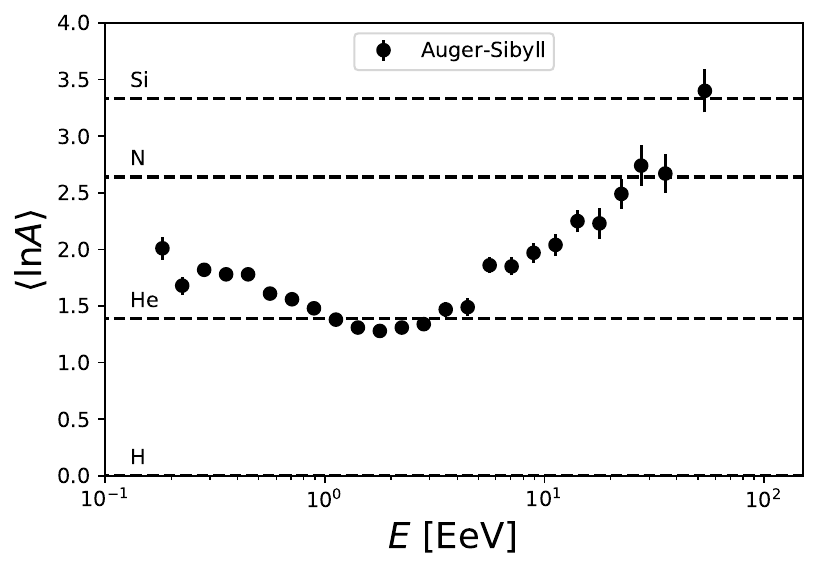}\includegraphics[width=0.3\textwidth]{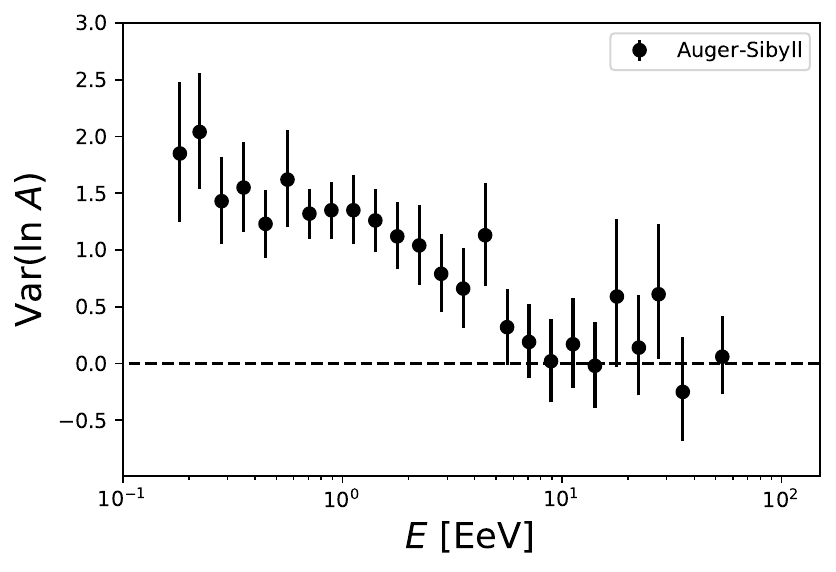}

    \caption{Measured spectrum \cite{ab21} and inferred $\langle{\rm ln}A\rangle$ and Var(ln$A$) for the Sibyll~2.3c hadronic model from \cite{yu19}, obtained with the Pierre Auger Observatory.}
    \label{f1}
\end{figure}

In Fig.~\ref{f1} we show the observables that  will be considered to fit our model, which are the measurement by the Pierre Auger Observatory of the spectrum (left panel, from \cite{ab21}) and the values of $\langle {\rm ln}A\rangle$ and of Var(ln$A$) inferred  on the basis of Sibyll~2.3c (central and right panels, from \cite{yu19}). Note that this experiment largely dominates the present statistics of UHECR observations, so that for simplicity we do not include other measurements. The Auger Observatory spectrum includes events up to an energy of about 160\,EeV, and although events with larger assigned energies were observed by the Telescope Array and Fly' s Eye experiments, one should keep in mind that they also involve different energy calibrations. 

By combining the spectrum and composition results one may obtain some important pieces of information. In particular, between 1 and 4~EeV one has that the  composition should on average be not heavier than He and, moreover, a minimum average ln$A$ is achieved near 2~EeV. The values of Var(ln$A$) progressively decrease for increasing energies and above the ankle energy they are consistent with an almost pure composition at any given energy (corresponding to Var(ln$A)=0$). 

We will consider that the emitted spectra can be described as power-laws with  rigidity dependent cutoffs, both for the nearby source dominating the flux above a few EeV and for the extragalactic population dominating at lower energies,  so that
 \begin{equation}
     \frac{{\rm d}\Phi_\alpha}{{\rm d}E}=\phi_0^\alpha\sum_if_i^\alpha \left(\frac{E}{E_0}\right)^{-\gamma_\alpha}F_{\rm cut}\left(\frac{E}{Z_iR_{\rm cut}^\alpha}\right),
 \end{equation}
 with the index $\alpha$ identifying the population considered,  with $\alpha={\rm L}$ for the  LE population or  or $\alpha=s$  for the nearby source (and we will use  $\alpha={\rm H}$ when  discussing a continuous distribution of sources for the HE population). The total differential flux of the population $\alpha$ at the reference energy $E_0$ (smaller than the proton cutoff energy $R_{\rm cut}^\alpha$, which will be  referred to as the rigidity cutoff\footnote{The actual magnetic rigidity is $R\equiv pc/Ze$, which in the relativistic limit is proportional to $E/Z$.}) is $\phi_0^\alpha$, and $f_i^\alpha$ are the fractions of the different elements $i$ of charge $Z_i$ emitted at the source at any fixed energy $E\ll R_{\rm cut}^\alpha$.  
 The cutoff function will be taken as
 \begin{equation}
     F_{\rm cut}(x)={\rm sech}(x^\Delta),
 \end{equation}
 with the parameter $\Delta$ determining the steepness of the cutoff.
The different parameters involved
will be determined so as to reproduce the observed spectrum and composition shown in Fig.~\ref{f1}. 

Let us mention that the fractions $f_i$, representing the relative contributions to the flux at a given energy, can be related \cite{mo18} to the overall elemental number fractions $f_i^0$ present in the source medium before acceleration if one assumes that all elements were completely ionised  and the acceleration was rigidity dependent and gave rise to a power-law with index $\gamma$. This relation would be $f_i^0= f_iZ_i^{1-\gamma}/(\sum_i f_i Z_i^{1-\gamma})$. In particular, for $\gamma\simeq 2$ the original fractions $f_i^0$ of the heavier elements in the medium will be suppressed with respect to those associated to the spectrum at a given energy, $f_i$, while for very hard spectra ($\gamma<1$) they would be enhanced. 

When the nuclei propagate from the sources through the CMB  photon background, whose present temperature is $T_{\rm CMB}\simeq 2.7$K, they can suffer photodisintegrations involving the emission of nucleons or $\alpha$ particles (nuclear fission is only relevant for nuclei heavier than Fe, whose contribution  we assume to be negligible).
The average energy of the background photons
is about $\bar{E}_\gamma \simeq 2.7T_{\rm CMB}\simeq 1$~meV. For an UHECR nucleus of energy $E$, mass number $A$ and Lorentz factor $\Gamma=E/(Am_p)\simeq 10^9E/(A\,{\rm EeV})$, when seen in the rest frame of the UHECR the photon energy will be $\varepsilon\simeq \Gamma E_\gamma(1-\cos\beta)$, with $\beta$ the angle between the cosmic ray  and the CMB photon trajectories. This energy is maximal for head-on collisions, in which case one has that $\varepsilon\simeq 2\,{\rm MeV}\,E/(A\,{\rm EeV})$. On the other hand, the threshold for the photodisintegration reaction is typically of order $\varepsilon_{\rm th}\simeq 10$~MeV (and about 20~MeV for the  more tightly bound nuclei such as He) \cite{st09}, with the peak of the disintegration cross section being at about 20 to 30~MeV, at the so-called  giant dipole resonance.  
For heavy nuclei  the photodisintegration proceeds mostly  by single or double nucleon emission, with the emission of $\alpha$ particles being subdominant (amounting typically to less than 1\% if energies well above the threshold contribute). However,   
when the disintegration chain reaches the $^9$Be nucleus this one quickly photodisintegrates, with a very low photon energy threshold of just 1.7~MeV, and produces two alpha particles and a neutron, so that this  process  provides the main contribution to the fluxes of secondary He nuclei.

\begin{figure}[t]
    \centering
    \includegraphics[width=0.33\textwidth]{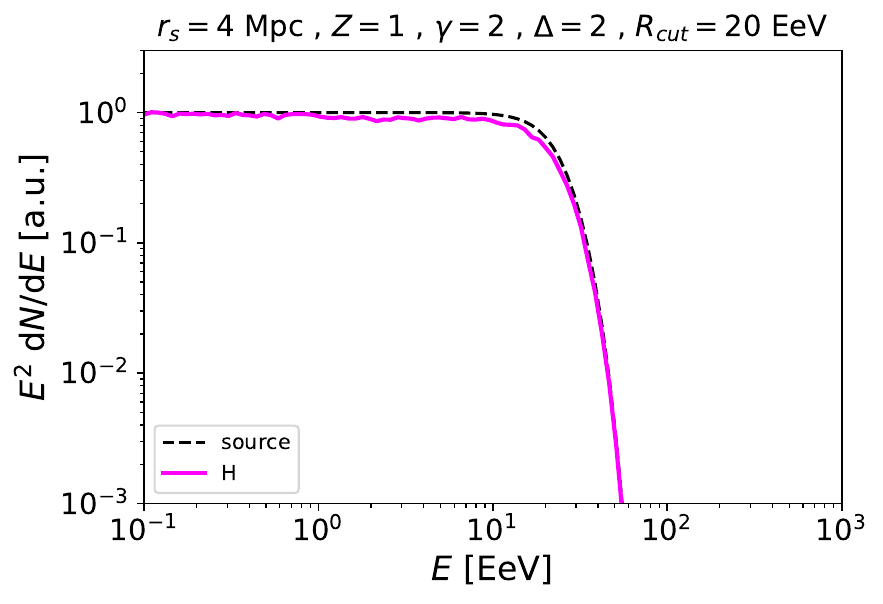}\includegraphics[width=0.33\textwidth]{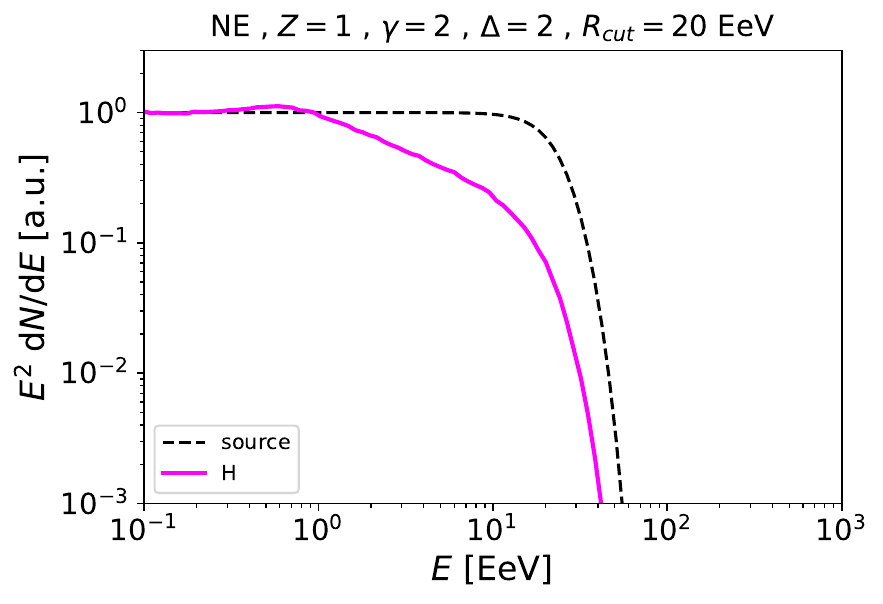}\includegraphics[width=0.33\textwidth]{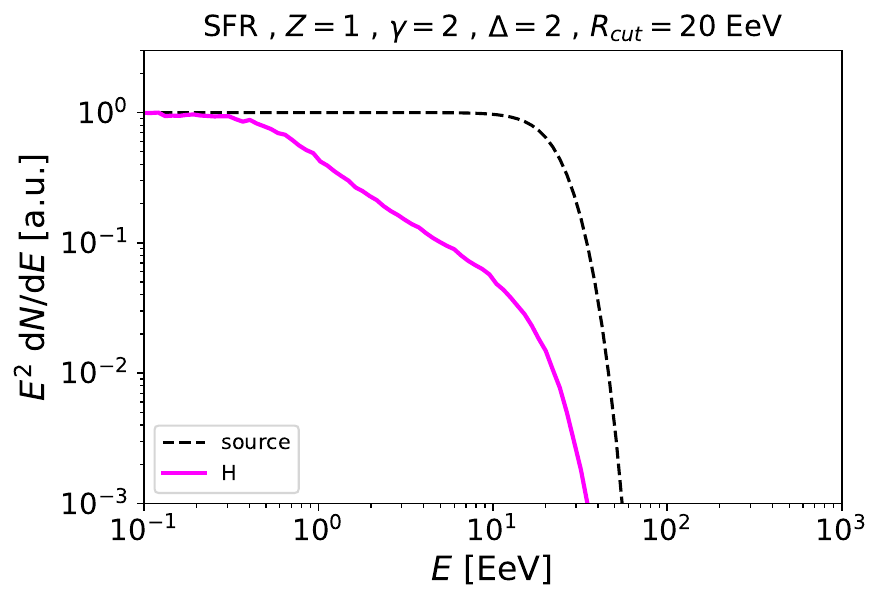}\\
\includegraphics[width=0.33\textwidth]{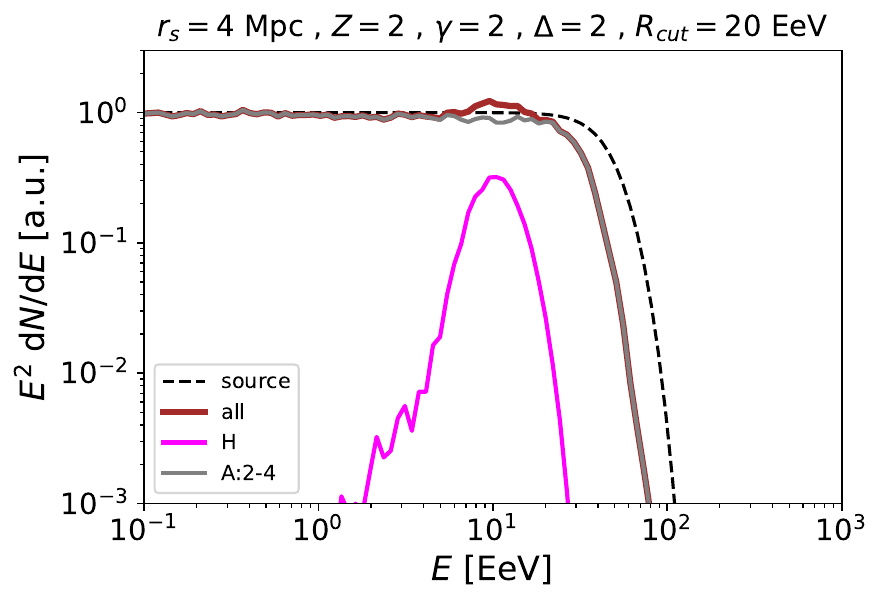}\includegraphics[width=0.33\textwidth]{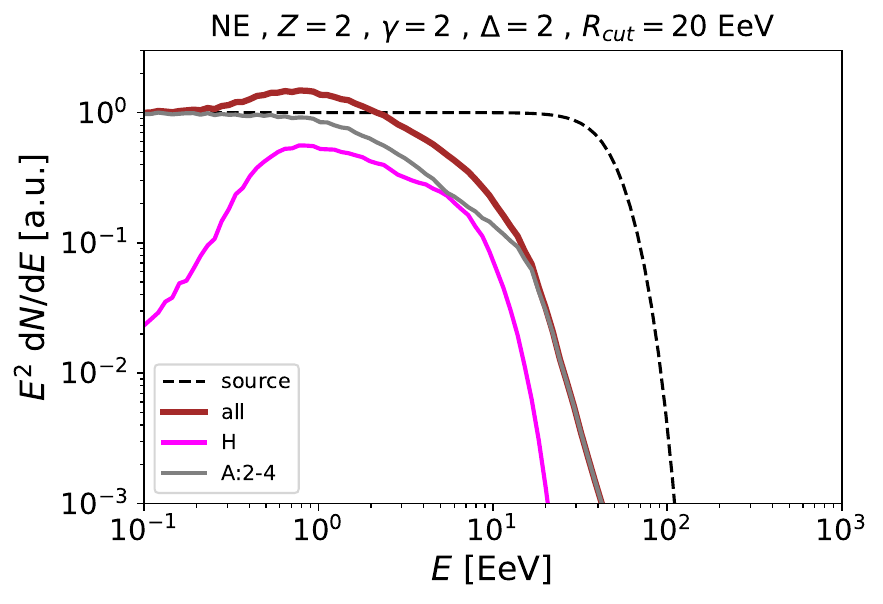}\includegraphics[width=0.33\textwidth]{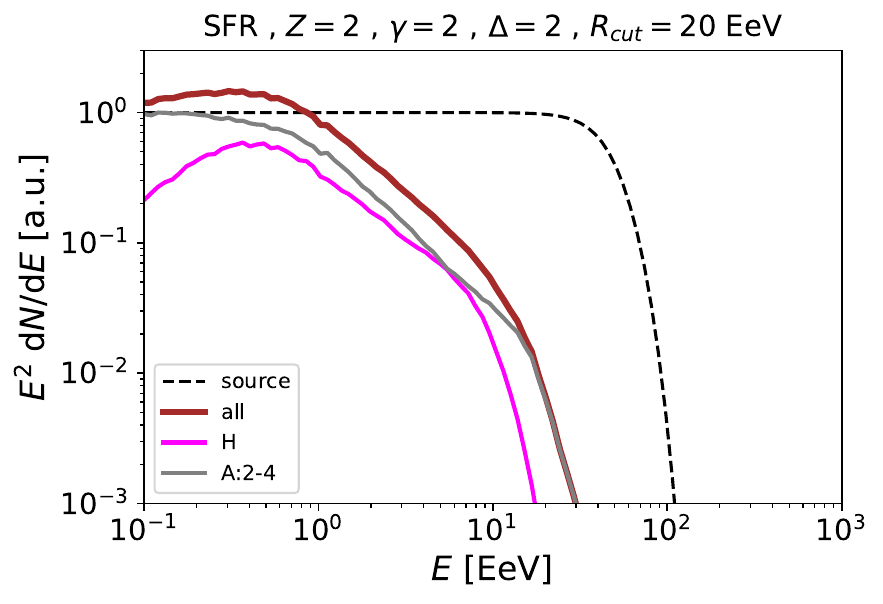}\\
     \includegraphics[width=0.33\textwidth]{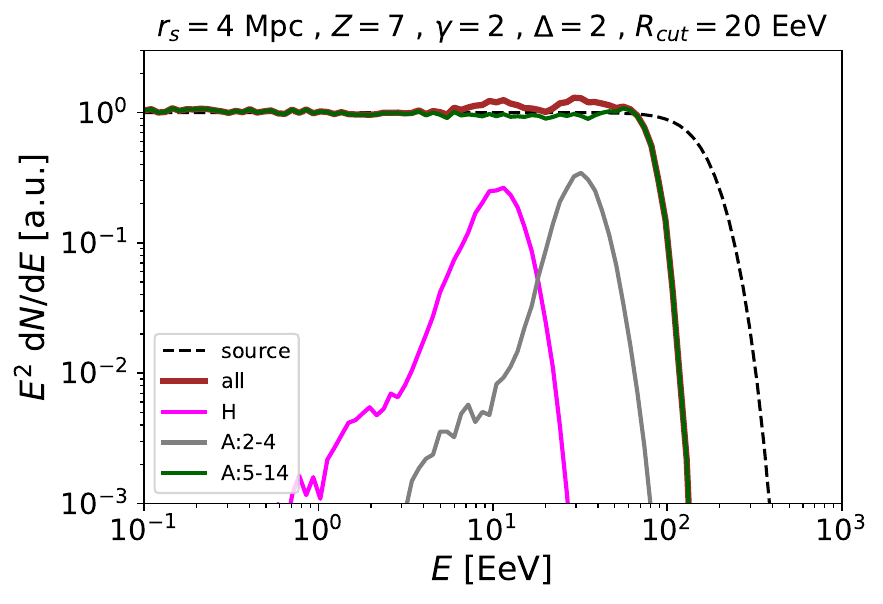}\includegraphics[width=0.33\textwidth]{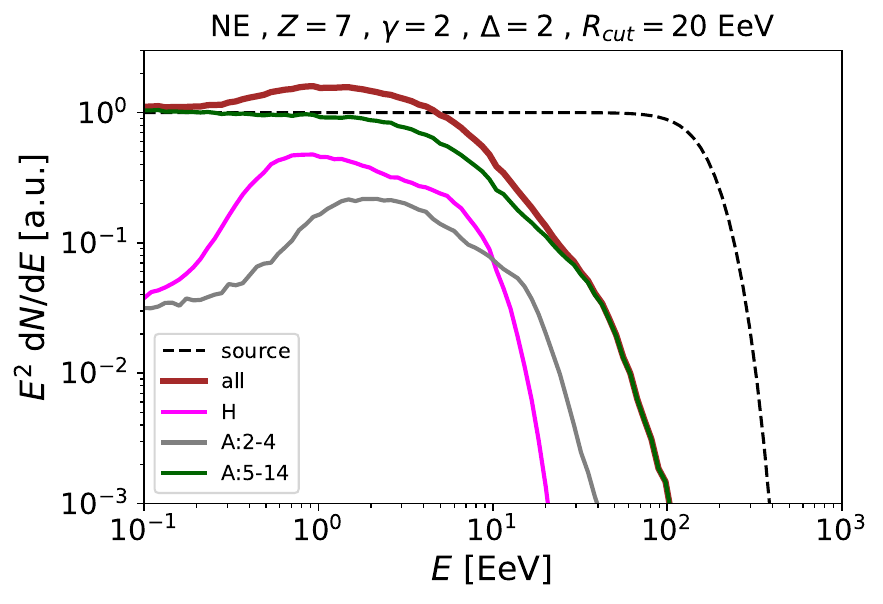}\includegraphics[width=0.33\textwidth]{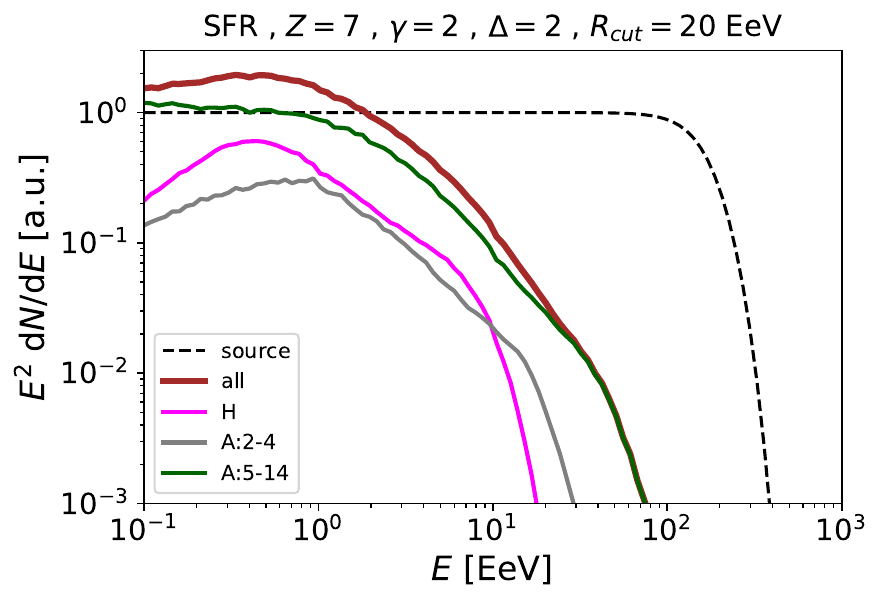}\\
     \includegraphics[width=0.33\textwidth]{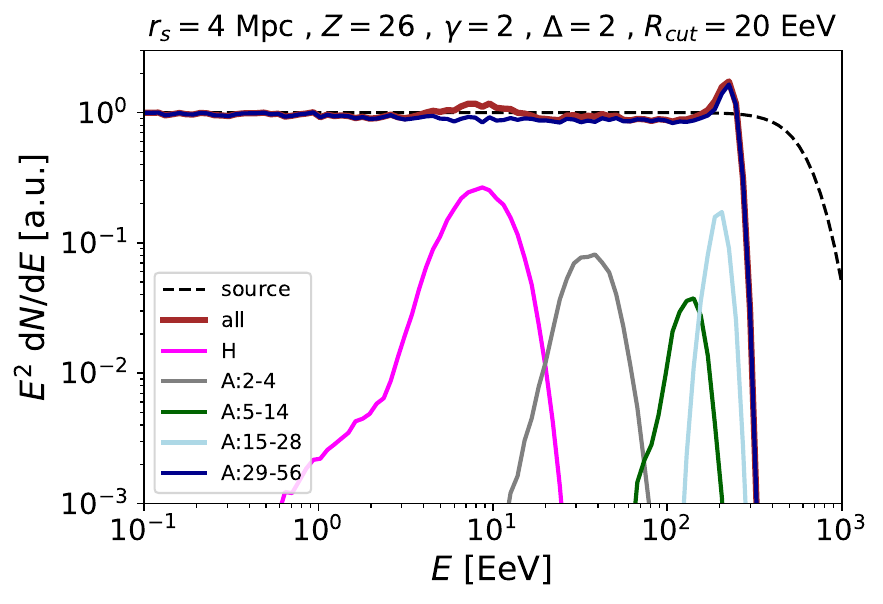}\includegraphics[width=0.33\textwidth]{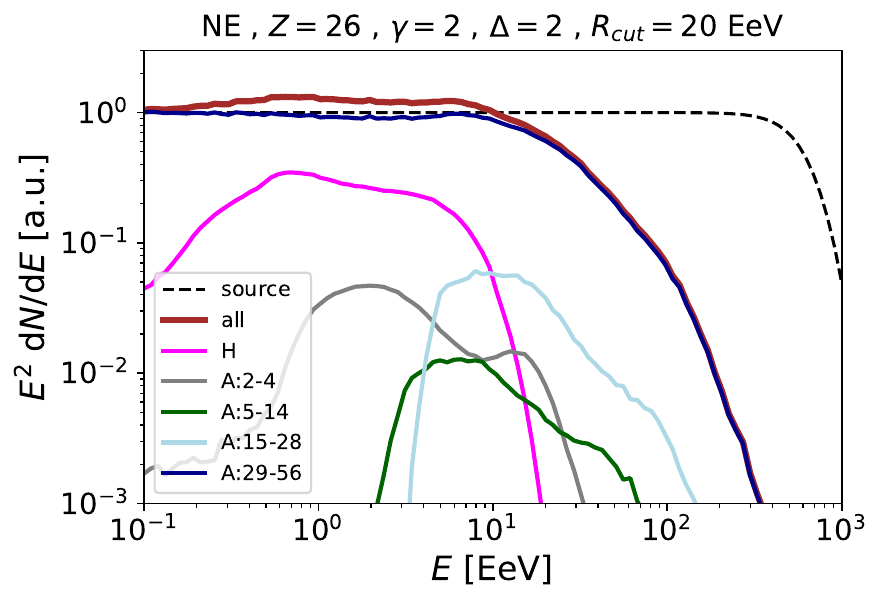}\includegraphics[width=0.33\textwidth]{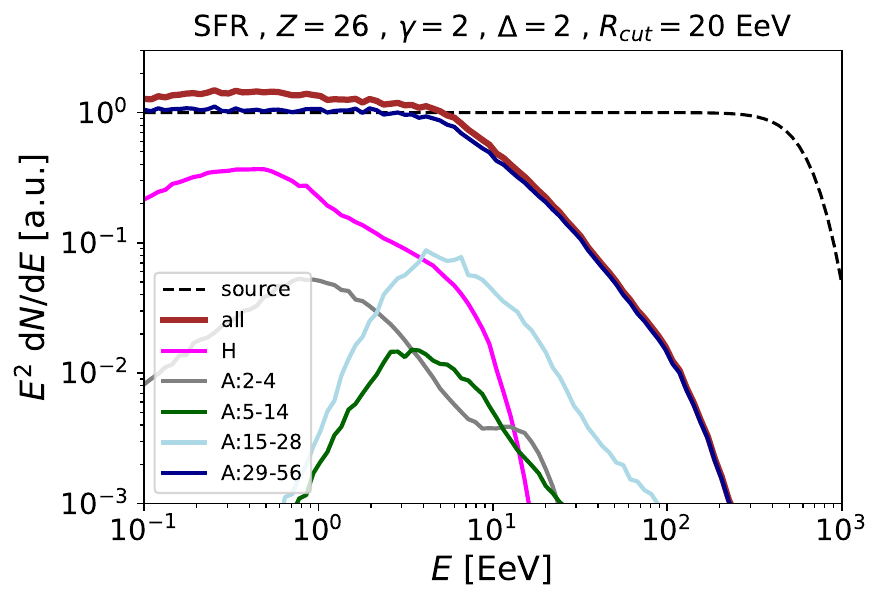}\\

    \caption{Fluxes of arriving CRs separated in different mass groups, for different injected elements (H, He, N and Fe from top to bottom) having a power-law spectrum with $\gamma=2$ and a rigidity dependent cutoff with $R_{\rm cut}=20$~EeV and $\Delta=2$ (dashed lines). From left to right are the cases of a nearby source at 4~Mpc distance, a homogeneous distribution of sources with NE or with SFR.}
    \label{f2}
\end{figure}

 In Fig.~\ref{f2} we illustrate the impact of the attenuation effects during propagation for different emitted primary CRs (H, He, N and Fe), adopting source spectra with $\gamma=2$ and considering a rigidity cutoff characterized by $R_{\rm cut}=20$~EeV and $\Delta=2$ (whose shape is shown with dashed lines). The results are obtained with the TALYS photodisintegration  cross sections \cite{ko23} and the EBL model from Gilmore et al. \cite{gi12}, using the SimProp code \cite{al17}. We display the results for different source populations, such as a single nearby source at a distance $r_{\rm s}=4$~Mpc (similar to the distance to Cen~A), a continuous distribution of sources up to a maximum redshift $z=1$ with no evolution  (indicated as NE) or one following the redshift evolution of the star formation rate up to a maximum redshift of 4 (SFR, adopting the detailed source evolution from \cite{ho06}).

\begin{figure}[t]
    \centering
\includegraphics[width=0.4\textwidth]{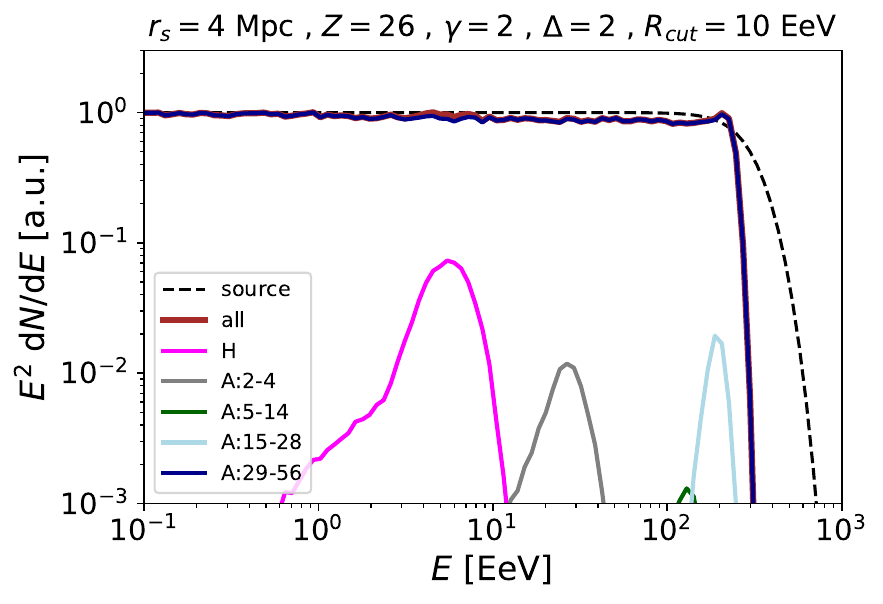}\includegraphics[width=0.4\textwidth]{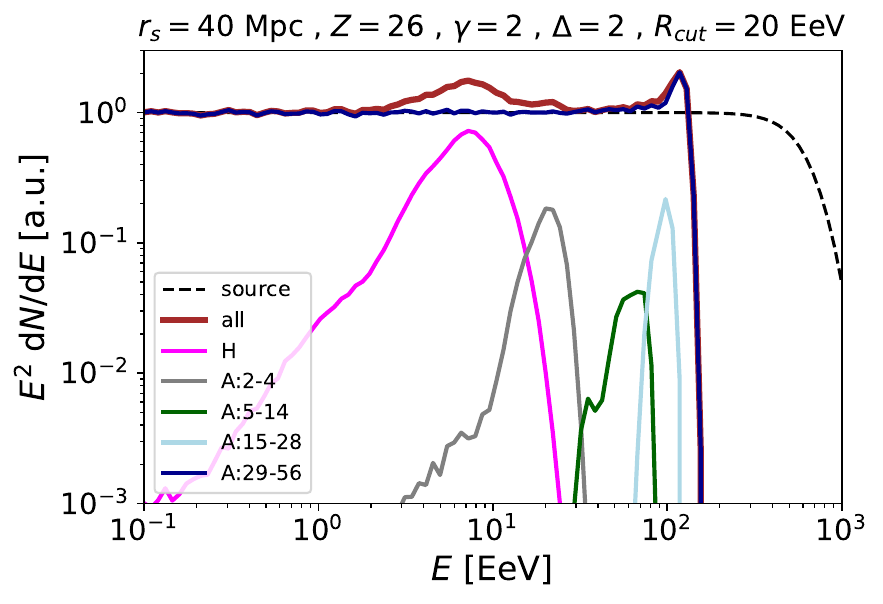}\\

    \caption{Fluxes of arriving CRs separated in different mass groups, for Fe nuclei injected  with $\gamma=2$  and $\Delta=2$. Left panel is for  a  source at 4~Mpc distance with a cutoff rigidity $R_{\rm cut}=10$~EeV while right panel  for  a  source at 40~Mpc distance with a cutoff rigidity $R_{\rm cut}=20$~EeV.}
    \label{f3}
\end{figure}

From the different panels on the left, associated to the nearby source case,  one can conclude that  the H component is not affected by the interactions, given the short source distance of just 4~Mpc. On the other hand, for heavier nuclei the photodisintegrations with CMB photons, whose threshold appears at a CR energy of about $5A$~EeV (and a bit higher for the tightly bound He nuclei), lead to a strong suppression of the arriving nuclei above this energy. This is apparent in the pronounced suppression of He nuclei observed above about 30~EeV, that of N nuclei above 70~EeV and that of Fe nuclei above about 250~EeV. One should note that the mean free path for photodisintegrations with the CMB at the giant dipole resonance peak, which takes place at $E\simeq 10A$~EeV, is curiously about $\lambda_{\rm mfp}\simeq 4\,{\rm Mpc}/A$, so that about $A$ interactions take place at these energies for nuclei of mass $A$ emitted from a source at 4~Mpc. One then expects an almost complete disintegration of those nuclei near the resonance energy, as is indeed observed in the figures corresponding to the nearby source.\footnote{One should also keep in mind that the Lorentz factor of the leading nuclear fragment is essentially unchanged in the interaction, and hence as long as the travelled distance is  larger than the interaction length  the cosmic rays will continue  to disintegrate along their trip, with the surviving nuclei becoming progressively lighter as the distance from the source  increases. } 
Due to the effect of the peak of the giant resonance, the H and He secondaries from the heavier nuclei show bumps peaking at energies near 10 and 30 EeV respectively, which also extend to lower energies due to the interactions with the Wien tail of the photon distribution. The heights of those peaks actually depend on the value of the cutoff rigidity considered. Cutoff values larger than 20~EeV would enhance the amount of secondaries with respect to those shown in the plots and would  also lead to secondary H extending up to higher energies, of order $R_{\rm cut}Z/A\simeq R_{\rm cut}/2$, with the shape of the suppression  approximately following the shape of the cutoff suppression of the primaries. On the other hand,  values of $R_{\rm cut}$ smaller than 20~EeV would reduce the height of those peaks, with the amount of secondaries becoming quite small when $R_{\rm cut}<10$~EeV (which will often turn out to be relevant in the present scenario). This is shown in the left panel of Fig.~\ref{f3} for the case of injected Fe nuclei. The main effect in this case is just that of chopping the Fe component beyond about 250~EeV, as a consequence of the emission of few nucleons during the propagation which end up with energies of about $R_{\rm cut}/2\simeq 5$\,EeV. In the right panel of Fig.~\ref{f3} we show instead the spectrum for injected Fe nuclei from a source lying at a distance of 40~Mpc, as could be for instance the case of the active galaxy NGC~4696 in the Centaurus cluster. In this case the cutoff at Earth due to interactions  will shift to lower energies, since interactions with the Wien tail of the photon distribution become more significant, and it would already be close to the highest energies observed with Auger of about 160~EeV (and smaller than the highest energies determined by the Telescope Array \cite{ab24b}). It is clear that the closest CR source cannot be much farther than 40~Mpc unless the emitted nuclei at the highest energies were much heavier than iron. Some constraints on the maximum distance to the closest source from a distribution of sources  were obtained e.g. in  refs.~\cite{ta11,la20} from the observed shape of the spectrum. 

For steeper spectra ($\gamma>2$) the fraction of secondaries would be smaller than the one considered in the plots, where $\gamma=2$ was adopted,  while they would be relatively more important for harder spectra ($\gamma<2$).

Regarding the results for a continuous source distribution, either with NE (middle panels of Fig.~\ref{f2}) or with SFR evolution (right  panels of Fig.~\ref{f2}),  some ingredients become important when the sources extend up to high redshifts. In particular, one  needs to take into account the redshift evolution of the photon densities and of the CMB temperature, as well as  the cosmological source evolution. Moreover, the more energetic EBL radiation also affects the nuclear disintegrations in a sizeable way down to lower CR energies.  For interactions with the CMB, the peak of the dipole resonance is achieved for nuclear energies of about $10A/(1+z)$~EeV, with $z$ the redshift at which the interaction takes place, and where we accounted for the increased CMB photon temperature at that redshift. The typical energy of the arriving redshifted nucleon fragments will then be $10/(1+z)^2$~EeV, and  will hence be distributed between several and 10~EeV if the CR sources contribute up to cosmological distances. The interactions with the more energetic EBL photons can produce large amounts of less energetic secondaries, typically extending down to energies   slightly below 1~EeV. On the other hand, secondary He nuclei would also peak at a few EeV, since for higher energies they will be strongly disintegrated if they come from far away. Only for very hard spectra, $\gamma<1$, the  secondary He would be prominent at higher energies  before they get strongly suppressed at $E>20$~EeV by interactions with the CMB. 
Note  that the spectra of the heavy nuclei start to steepen, due to the photodisintegrations with the EBL, at energies of about $0.2A$~EeV in the NE case and at about $0.1A$~EeV in the SFR case. Also the primary H component from a population of cosmologically distributed sources steepens  above about 1~EeV due to the opening of the threshold for pair production off CMB photons.

From these results one may also see that a plausible explanation for the minimum average mass observed near 2~EeV is that it be the consequence of the presence of significant amounts of secondary H (and He) nuclei produced in the photodisintegration of heavier nuclei interacting with the CMB and EBL as they propagate from faraway sources. 
The increase of $\langle {\rm ln}A\rangle$ below 2~EeV  may be due to a combination of the reduced amount of light secondaries for decreasing energies, leading to an increase in the proportion of CRs from the heavier primaries of the LE population, eventually also enhanced by the presence of the tail of the Galactic contribution of heavy nuclei. 

In the scenario with a nearby source dominating above the ankle energy,  the production of secondaries from the nearby source is moderate and they   appear at higher energies, hence the light secondary particles eventually appearing at a few EeV should be mostly due to the LE component.

\section{Reproducing the observed spectrum and composition}
We will focus  in this section on the case with no magnetic fields and with steady sources, addressing in the next section the effects of extragalactic and Galactic magnetic fields, as well as the impact of a finite source lifetime.

\subsection{The baseline scenarios}
With the model described before, consisting of a nearby source at 4~Mpc distance dominating the flux at the highest energies and a continuous distribution of sources becoming important below the ankle energy, we searched for the  spectral and composition parameters leading to the best agreement with the observations above $10^{17.8}\ {\rm eV}\simeq 0.63$~EeV (at lower energies  the LE component may not have a simple power-law source spectrum and  also additional source populations could  become relevant).  We will first focus on the assumption that just those two populations are present and that they emit H, He, N, Si and Fe nuclei. For definiteness we will concentrate first on the case with a cutoff shape having $\Delta=2$, for which the spectral index and rigidity cutoff have a more clear interpretation, and  later on discuss also the case with $\Delta=1$ which involves a broader cutoff suppression. Note that since  the nearby source dominates the flux above the ankle energy and in this section we are ignoring the magnetic field deflections,  this scenario is not expected to reproduce the observed arrival directions.

\begin{figure}[t]
    \centering
    \includegraphics[width=0.4\textwidth]{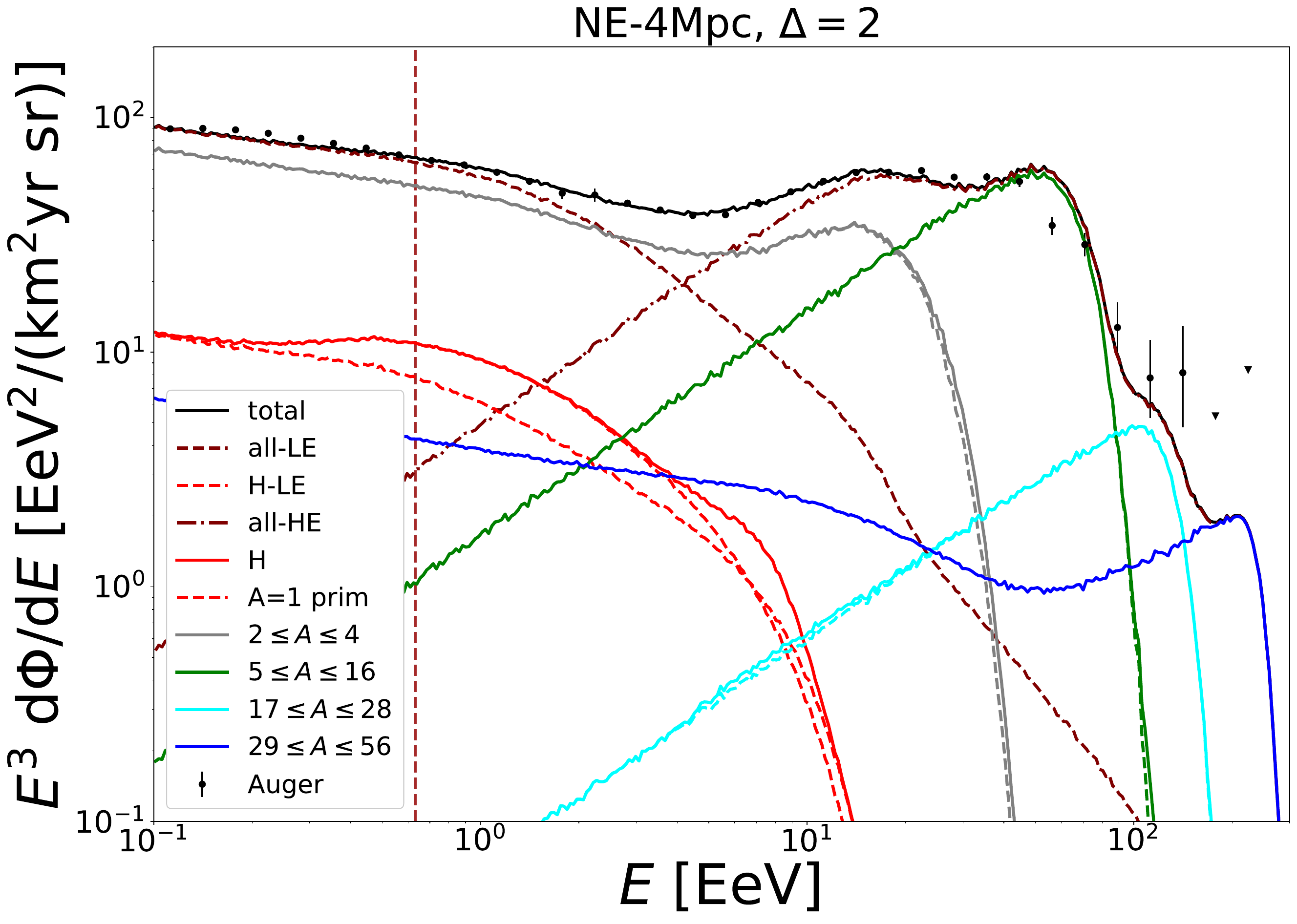}\includegraphics[width=0.4\textwidth]{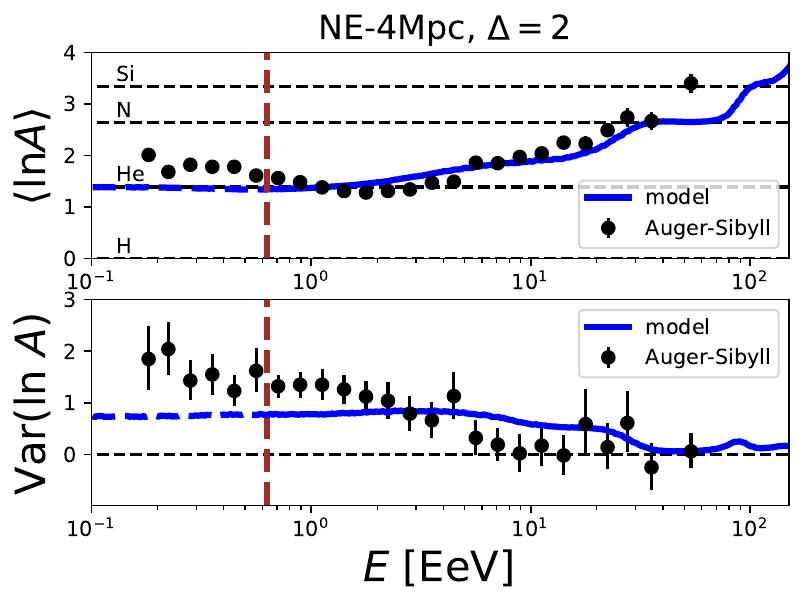}\\
    \includegraphics[width=0.4\textwidth]{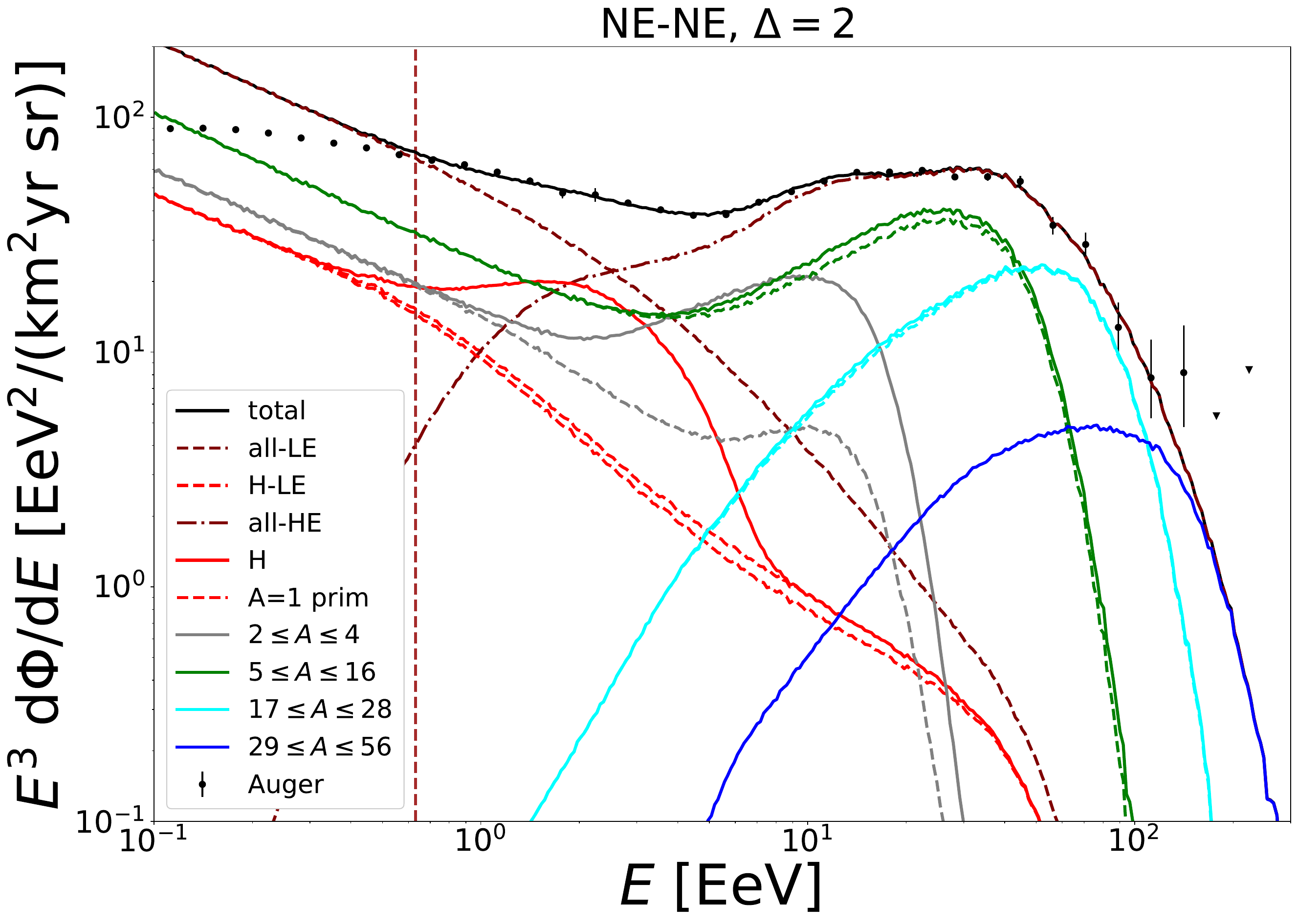}\includegraphics[width=0.4\textwidth]{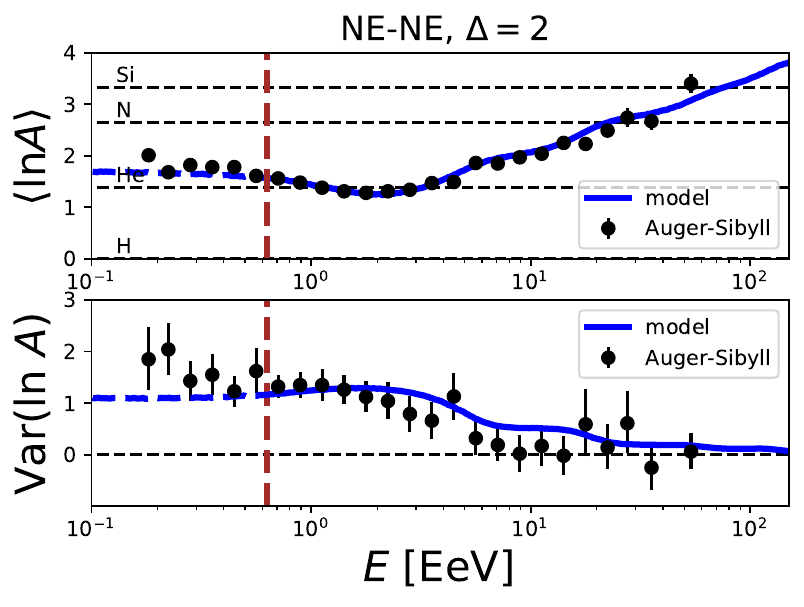}\\

    \caption{Left panels:  Total measured spectrum as well as the different fitted  components. Right panels: Values of $\langle {\rm ln}A\rangle$ and Var(ln$A$) inferred from $X_{\rm max}$ data using Sibyll~2.3c (dots) as well as those obtained in the fit (blue lines). Vertical dashed lines indicate the threshold of the fit. The plots consider $\Delta=2$, NE for the low-energy population and for the high-energy population a nearby source at 4~Mpc (top panels) or a continuous distribution with NE (bottom panels).}
    \label{fD2noG}
\end{figure}

In Fig.~\ref{fD2noG} we display (top panels) the results of the  fit with the nearby source at 4~Mpc distance and considering that the sources of the low-energy population have no cosmological evolution (scenario NE--4Mpc).  For reference, we also include (lower panels) the results that would be obtained if instead of a single nearby source there was a different population of sources distributed uniformly and with no-evolution (scenario NE--NE), similar to some of the scenarios  considered in \cite{ab23}. One peculiar difference in this case is the shape of the suppression at the highest energies observed, with some recovery due to the Fe component being apparent in the case of the nearby source. Let us note that the spectral shape at the highest energies should depend sensitively on the relative Si and Fe abundance ratio, which is actually not strongly constrained at present given the scarce information on the composition in this energy range (the highest energy bin for the composition corresponds to $E> 10^{19.6}\,{\rm eV}\simeq 40$~EeV). Hence, more detailed studies of the composition with the increased statistics  of the Auger Observatory surface detectors \cite{aa16} and a detailed study of the spectral shape at the highest energies should help to further constrain these scenarios.
Also note that in the NE-NE scenario, below the threshold of the fit the spectrum tends  to  overshoot the measurements, so that in this case the assumption of a very steep power law extending to lower energies would need to be reconsidered, or magnetic horizon effects accounted for, or a stronger source evolution adopted, or a Galactic component included \cite{mo20}.

We report in the first four rows of Table~\ref{tD2noB} the spectral and composition  parameters obtained in the fits of the different scenarios considered. We also include here  the cases in which the low-energy population has an evolution following the SFR  (scenario SFR-4Mpc and SFR-NE). The main change in these last cases is that due to the stronger attenuation effects present when the SFR is considered, a harder spectrum for the LE population is inferred ($\gamma_{\rm L}$ smaller by about 0.3 to 0.5). No major changes in the spectral parameters of the high-energy population result in these cases, and we will hence for simplicity not discuss them further.\footnote{In the case of two population of sources, the slight differences in the fitted parameters with respect to \cite{ab23}  are related to the hadronic model considered, the maximum redshift adopted and the fact that we fit the mean and variance of ln$A$ rather than the whole $X_{\rm max}$ distributions.}  We also display in the table the overall $\chi^2/{\rm dof}$ of the fits. Taking into account that there are 26 spectral energy bins (two of them containing  just upper bounds) and 19 composition energy bins, for a total of $N=64$ data-points, while the number of fitted parameters is just 14 (for each population there is the flux normalisation, spectral index and rigidity cutoff, together with the four independent elemental fractions), the values of the $\chi^2$ obtained are not unreasonable given the simplified nature of the models considered to describe a certainly more complex reality, and possible improvements in this direction will be discussed later on. Although  the continuous source distribution models have a  smaller value of $\chi^2/$dof than the baseline Cen~A model considered in this section, we will see that the different refinements of this last model discussed in the following sections will bring it to a similar or better overall fit quality.

\begin{table}[ht]
\addtolength{\tabcolsep}{-1pt}
\centering

{\small
\begin{tabular}[H]{ @{}c| c  c c c c c c| c c c c c| c @{}}
\multicolumn{14}{c}{ no B,  $\Delta=2$}\\
\hline
model & $\gamma_{\rm H}$ & $R_{\rm cut}^{\rm H}$ & $f^{\rm H}_{\rm H}$ & $f^{\rm H}_{\rm He}$ & $f^{\rm H}_{\rm N}$ & $f^{\rm H}_{\rm Si}$ & $f^{\rm H}_{\rm Fe}$ &$\gamma_{\rm L}$ & $R_{\rm cut}^{\rm L}$ & $f^{\rm L}_{\rm H}$ & $f^{\rm L}_{\rm He}$ & $f^{\rm L}_{\rm N}$ &  $\frac{\chi^2}{\rm dof}$ \bigstrut[t]\\ 
 &  & [EeV] &  & & [\%] &  &  &  &  [EeV]  &  & [\%] & \bigstrut[b]\\
\hline

NE-4Mpc &  2.0 & 8.3 & 0.4 &  64 & 34 & 1.3 & 0.3 &3.2 & 8.9 & 13 & 80 & 0 & 7.76 \bigstrut[t]\\ 
SFR-4Mpc & 2.0 & 7.6 & 0  & 58 & 40 & 1.4 & 0.5 & 2.7 & 11 & 0 & 91 & 8.7 & 8.60  \\
NE-NE  & 0.69 &  6.1 &  0 &  39 & 53 & 6.9 & 0.4 & 3.6 & 56 & 22 & 28 & 50 & 3.16 \\
SFR-NE & 0.76 & 6.3 & 12 & 32 & 48 & 6.8 & 0.4 & 3.3 & 64 & 23 & 32 & 45 & 3.38 \\
\hline
G-NE-4Mpc & 2.0 & 7.9 & 0 & 64 & 35 & 1.3 & 0.4 & 3.0 & 5.7 & 23 & 73 & 0 & 6.15 \bigstrut[t]\\
G-NE-NE & 0.73& 6.0 & 0 & 27 & 63 & 9.8 & 0.4 & 3.4 & 69 & 17 & 55 & 28 & 3.19 \\
\hline

\end{tabular}}

\caption{Parameters of the fit to the flux and composition for the different scenarios (see text).   When considering the nearby source one has that ${\rm H}=s$. For the LE population the Si fraction is negligible, and eventually the Fe component may provide the remainder of the other fractions. }
\label{tD2noB}
\end{table}

Some of the features which can be highlighted are:
\begin{itemize}
    \item For the adopted spectral shape with $\Delta=2$, the  inferred spectral index of the nearby source is close to two, which is in the ballpark of the expectations from diffusive shock acceleration, while for a continuous source distribution the spectrum turns out to be much harder, with $\gamma_{\rm H}<1$. This is related to the fact that no significant steepening of the spectrum of the nearby source gets produced below the cutoff value by the effects of propagation, unlike what happens for a population of sources extending up to high redshifts, which then require harder source spectra to counterbalance this effect. 

    \item For the case with  $\Delta=2$ the nearby source cutoff energy for nuclei of atomic number $Z$,  $E^{\rm s}_{\rm cut}=ZR^{\rm s}_{\rm cut}$, is a good indicator of the maximum energies achieved (while for $\Delta=1$ the falloff would be much broader). This maximum energy is found to be about $8Z$~EeV,  being slightly smaller than the energy $5A$~EeV  at which photodisintegrations of the nuclei at the giant dipole resonance start to become strong. Hence,  one can see that the main effect contributing to shape the suppression of the different elements turns out to be the source cutoff rather than the attenuation during propagation.  Given that CR nuclei are actually observed not far from the expected interaction cutoff, it is clear that the source cutoff cannot be much smaller than the interaction cutoff but, on the other hand, much larger values of the source cutoff would lead to too large amounts of secondaries being produced by interactions. In particular, secondary H nuclei would appear just around the ankle energy and secondary He nuclei extending somewhat beyond the instep feature, which would turn out to be problematic to fit the observations. These features then constrain the source cutoff to be not larger than the energy at which propagation effects would produce a strong suppression.

\item  The contribution of primary H nuclei from the nearby source turns out in general to be  negligibly small, but primary He nuclei provide a significant contribution in the instep region and N nuclei extend to higher energies up to the suppression, beyond which mostly Si and Fe nuclei survive, with the interactions during propagation producing some steepening at the end of the Fe component.

    \item  If the cutoff of the LE population is large, as happens in the NE-NE scenario, the light  components of the LE population extend well beyond the ankle energy. Hence,  due to the propagation effects the He component  will have  a cutoff close to the instep  and can help to account for this feature, which is anyhow found to be mostly due to the He component of the HE population since the spectrum of the LE population is quite steep.

\item In comparison with the scenarios with an additional continuous source distribution dominating above the ankle, in which large amounts of secondary He from N disintegrations appear  between  a few EeV and the instep and secondary H appear between 0.5~EeV and $R_{\rm cut}^{\rm H}/2$, no significant amounts of secondary H and He are produced by the nearby source, and hence the LE population has to provide most of the light nuclei below the ankle. In the scenario with the nearby source a too small variance of ln$A$ results at EeV energies, since below the ankle  the contribution from the nearby source becomes quite small while the LE population should consist mostly of He nuclei. This suggests that the tail of the Galactic component, which consists mostly of heavy Si and Fe group elements, could be beneficial to help to increase this variance, as we now discuss.

\end{itemize}

\subsection{Adding the Galactic CR contribution}

\begin{figure}[t]
    \centering
    \includegraphics[width=0.4\textwidth]{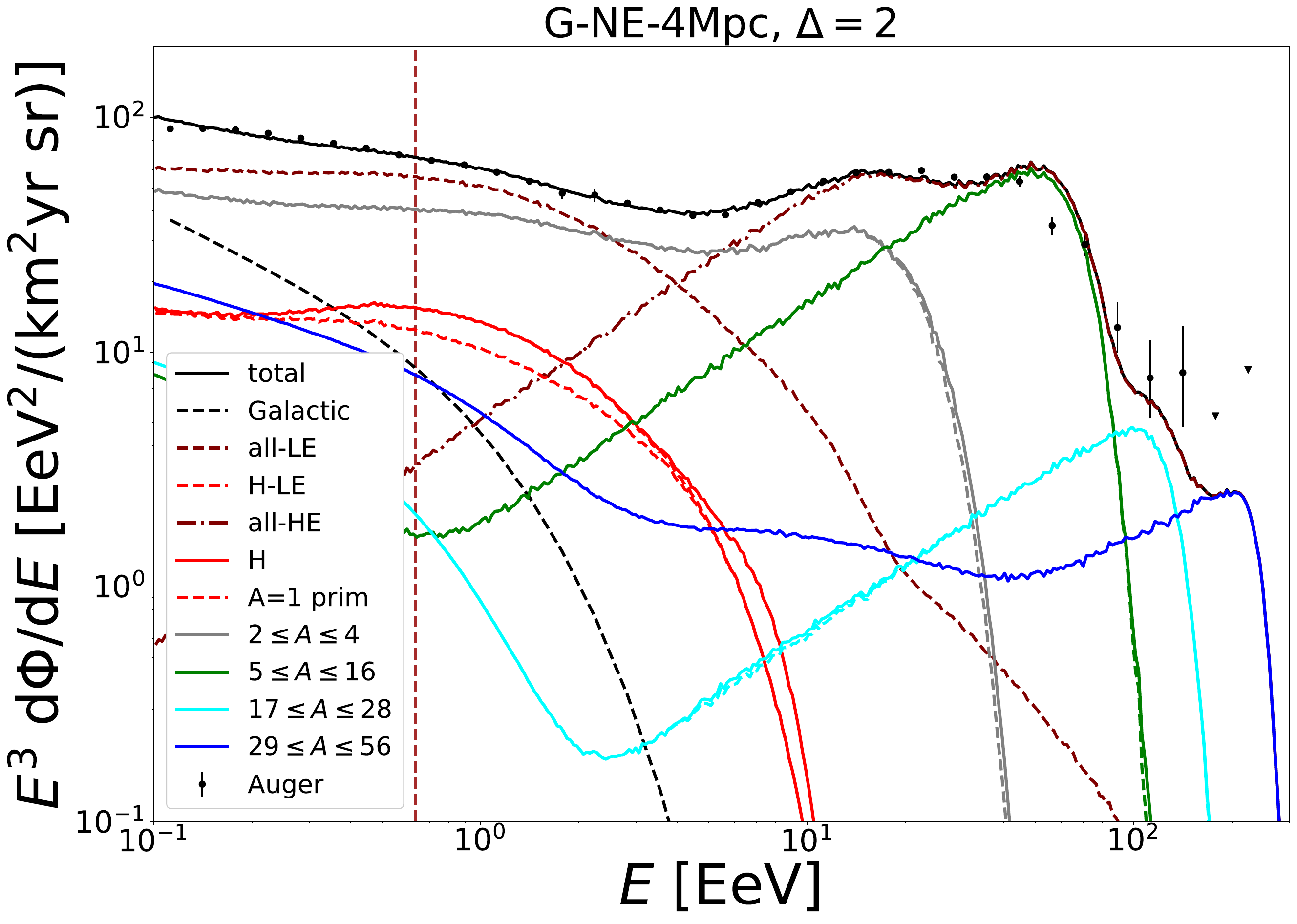}\includegraphics[width=0.4\textwidth]{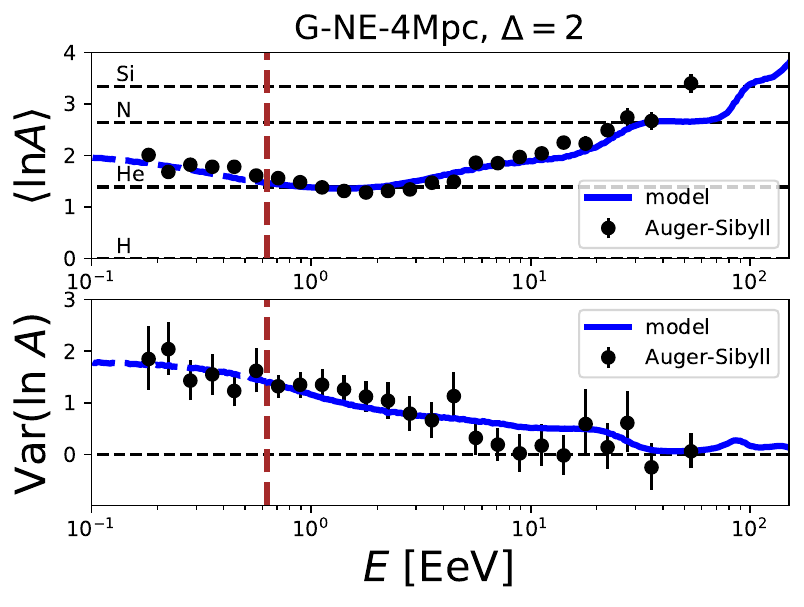}\\
    
     \includegraphics[width=0.4\textwidth]{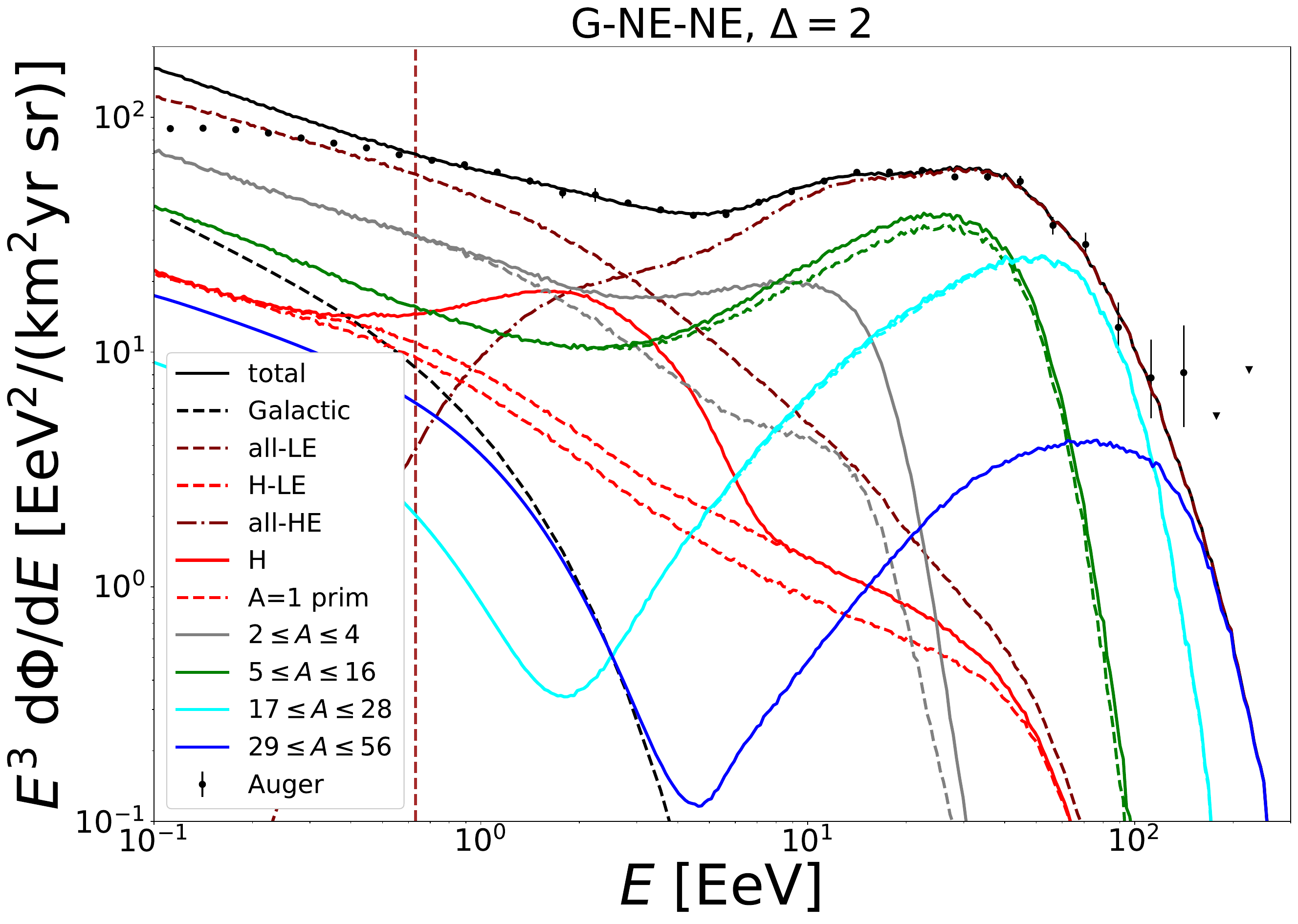}\includegraphics[width=0.4\textwidth]{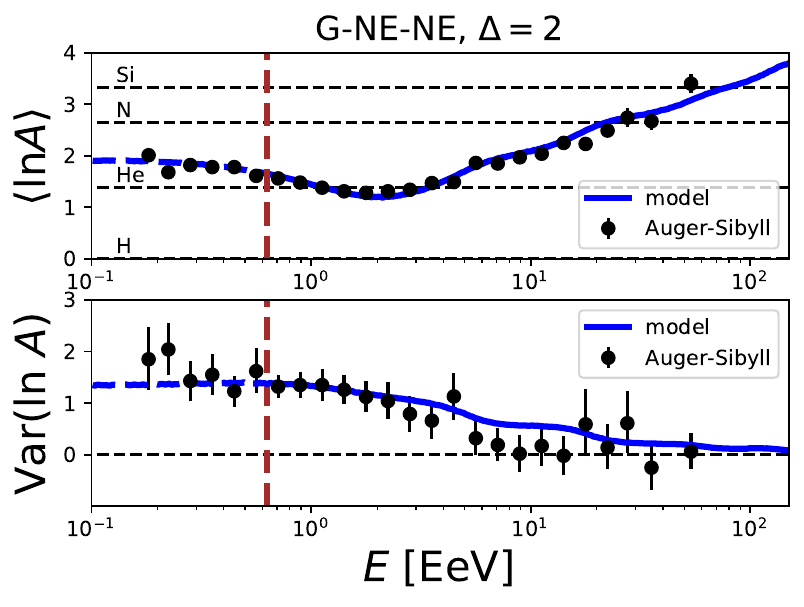}\\
   
    \caption{As in Fig.\,\ref{fD2noG} but for the scenarios including the Galactic component.}
    \label{fD2G}
\end{figure}

In order to understand the impact of the Galactic CR population, we include it here following the results obtained in \cite{mo19}, where measurements from various experiments at lower energies, covering the PeV to EeV range which encompasses  the knee and  the second knee features (at 4~PeV and 100~PeV respectively), where used to fit the Galactic CR spectrum at Earth. The same five representative elements were considered, and a smooth broken power-law rigidity dependent spectrum, with a high-energy cutoff, was obtained for each of them, of the form:
\begin{equation}
\frac{{\rm d}\Phi_{\rm G}}{{\rm d}E}=\phi_{\rm G}\sum_i\ f_i^{\rm G} \left(\frac{E}{\rm EeV}\right)^{-\gamma_1}\left[1+\left(\frac{E}{Z_iE_{\rm k}}\right)^{\Delta\gamma/w}\right]^{-w} \,{\rm sech}\left( \frac{E}{Z_i R_{\rm cut}^{\rm G}}\right),
\label{galflux} 
\end{equation}
with  parameters  $f_{\rm H}^{\rm G}=f_{\rm He}^{\rm G}=0.35$, $f_{\rm N}^{\rm G}=0.12$, $f_{\rm Si}^{\rm G}=0.08$ and $f_{\rm Fe}^{\rm G}=0.1$ while  $\gamma_1=2.76$, $\Delta\gamma=0.7$, $w=0.11$, $E_{\rm k}=a\,3.1$~PeV, $R_{\rm cut}^{\rm G}=a\,37$~PeV and $\phi_{\rm G}=a^{\gamma_1-1}\,425$~(km$^2$\,sr\,yr\,EeV)$^{-1}$  (see the fourth column of Table~3 in \cite{mo19}), and where we introduced the factor $a=1/1.07$ to account for the difference between the energy scales of the Telescope Array (adopted in \cite{mo19}) and that of the Auger Observatory (adopted here), and we note that the fluxes from the two experiments are normalized so as to approximately match at EeV energies.

The parameters obtained including this Galactic component are reported in the last two rows of Table\,\ref{tD2noB}, labelled as G-LE-HE with LE and HE indicating the source scenario considered for each population. The values for the G-NE-4Mpc and the G-NE-NE scenarios are given, and the resulting spectrum and composition are displayed in Fig.\,\ref{fD2G}.

The main salient points of including the Galactic component in the nearby source scenario are:
\begin{itemize}
    \item Since the Galactic component has a steep spectrum,  the spectrum of the LE population becomes harder.

    \item Given that the Galactic component is already quite heavy, the H contribution to the LE population needs to become larger. This ends up increasing the variance of ln$A$ at EeV energies, leading then to a better agreement with the observations than in the case with no Galactic population.

\item The  fit to the data improves when the Galactic component is introduced ($\chi^2/$dof gets reduced from 7.8 to 6.1).

\end{itemize}

Note that the heavy Galactic component that we consider does not contribute more than 10\% of the total flux at 1~EeV, and thus it should not lead to too large anisotropies at these energies, in agreement with observations.

 The main remaining problem with the fit considered  is that of not being  able to reproduce the details of the spectral suppression near 50~EeV, with N nuclei leading to a very pronounced bump near the suppression for  $\Delta=2$, while $\Delta=1$ would lead instead to a very broad one and intermediate values would anyhow also not be able to reproduce accurately the observed structure of the suppression. To tackle this point, we considered scenarios in which the HE population emits  C and O nuclei instead of just N ones.

\subsection{Injection of carbon and oxygen rather than nitrogen?}

\begin{table}[b]
\addtolength{\tabcolsep}{-1pt}
\centering
{\small
\begin{tabular}[H]{ @{}c| c  c c c c c c c| c c c c c| c @{}}
\multicolumn{15}{c}{model G-NE-4Mpc(CO), no B  }\\
\hline
$\Delta$ & $\gamma_{\rm s}$ & $R_{\rm cut}^{\rm s}$ & $f^{\rm s}_{\rm H}$ & $f^{\rm s}_{\rm He}$ & $f^{\rm s}_{\rm C}$ & $f^{\rm s}_{\rm O}$ & $f^{\rm s}_{\rm Si}$ & $f^{\rm s}_{\rm Fe}$ &$\gamma_{\rm L}$ & $R_{\rm cut}^{\rm L}$ & $f^{\rm L}_{\rm H}$ & $f^{\rm L}_{\rm He}$ & $f^{\rm L}_{\rm Fe}$ &  $\frac{\chi^2}{\rm dof}$\bigstrut[t] \\ 
 &  & [EeV] &  & & [\%] &  &  &  & &  [EeV]  &  & [\%] &  &\bigstrut[b]\\
\hline
2& 2.0 & 7.1 & 0 & 53 & 37 & 5.7 & 3.5 & 0.3 & 3.0 & 7.9 & 19 & 78 & 2.6 & 4.24 \bigstrut[t]\\
1& 1.4 & 1.8 & 0 & 30 & 52 & 8.7 & 8.9 & 0.6 & 3.0 & 50 & 12 & 87 & 1.5 & 3.89  \\
\hline

\end{tabular}}

\caption{Parameters of the fit to the flux and composition for the different scenarios considering emission of C and O from the nearby source rather than N (see text).  Results are shown for $\Delta=2$ and 1.
For the LE component the Si fraction is negligible and eventually the N provides the remainder fraction.}
\label{tGCO}
\end{table}

One basic motivation to consider scenarios injecting C and O is that it is indeed those elements which are found in nature to dominate  the CNO group, rather than the N, whose consideration is  just an effective way of describing this group of elements. Carbon nuclei get produced mostly through the triple alpha process in He burning stars, and O results mostly from the additional capture of an alpha particle by a C nucleus. Intermediate mass stars ($0.5\,M_\odot<M<8\,M_\odot$) actually end up their lives  as CO degenerate white dwarfs and the tidal disruption of these stars in the neighbourhood of intermediate mass black holes ($M_{\rm BH}\sim 10^3$ to $10^5M_\odot$), with the possible formation of powerful jets,  is a plausible way to inject those nuclei in an acceleration process that may reach ultrahigh energies \cite{zh17,ba17,bi18}. Moreover, another remarkable feature of the fits is the small amount  of light elements which is required to be injected in the high-energy population, and this could  be naturally understood  in this context.
Also more massive stars have large amounts of C and O elements outside their iron core before collapsing into a supernova. The supernova explosion, or also the tidal disruption of the star by a supermassive black hole ($M_{\rm BH} \gg 10^5M_\odot$) \cite{fa09} before the core collapse, could accelerate those  elements,
which 
in the scenario we consider should be further accelerated in the AGN jet to reach the highest CR energies. If the envelope of light elements of the star was expelled at earlier times, one could also understand the cause for the paucity of light elements in the HE population which is inferred from observations.  On the other hand, in a supernova explosion the Fe core should mostly end up (after getting photodisintegrated)  into the central compact remnant.

\begin{figure}[t]
    \centering
    \includegraphics[width=0.4\textwidth]{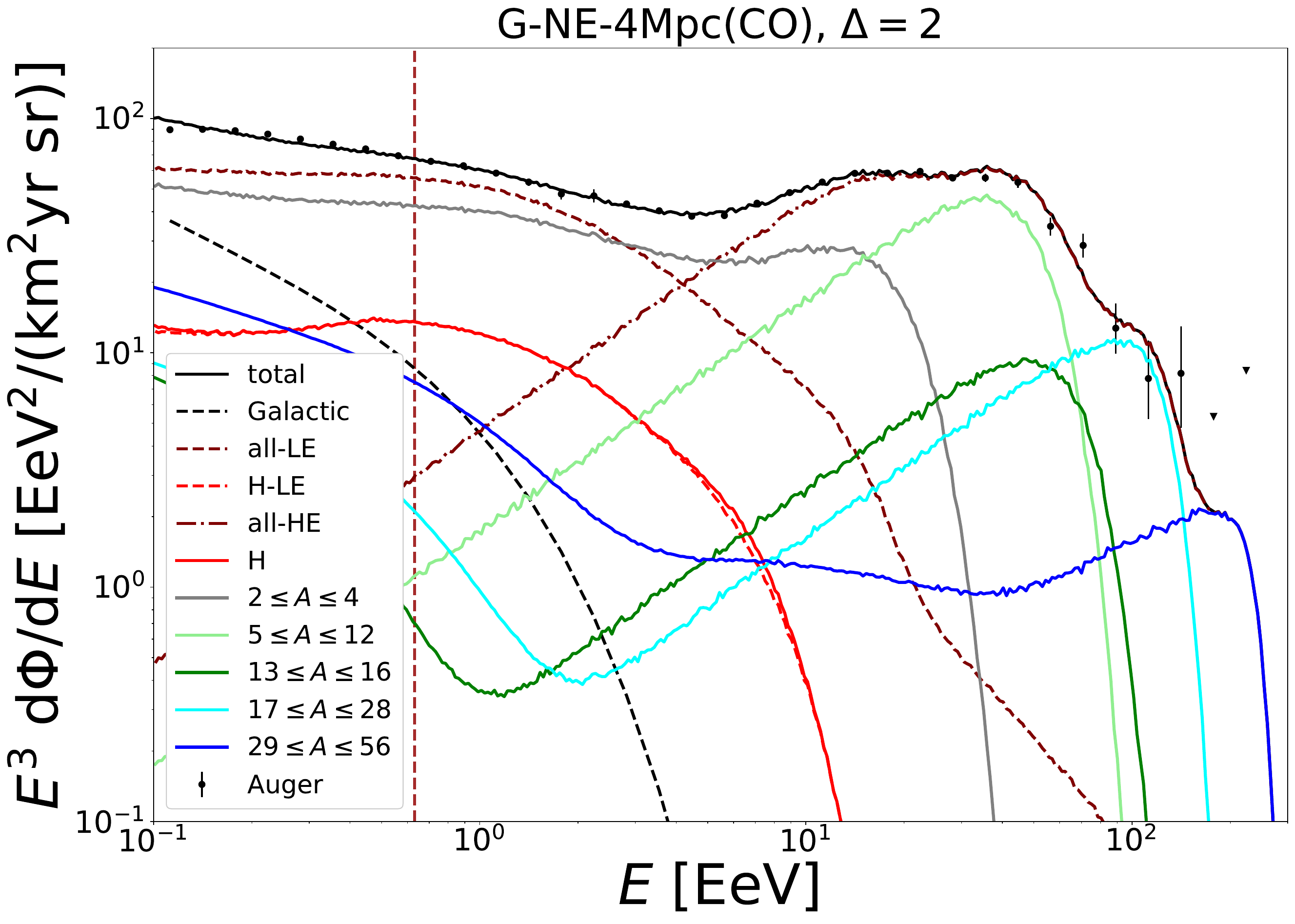}\includegraphics[width=0.4\textwidth]{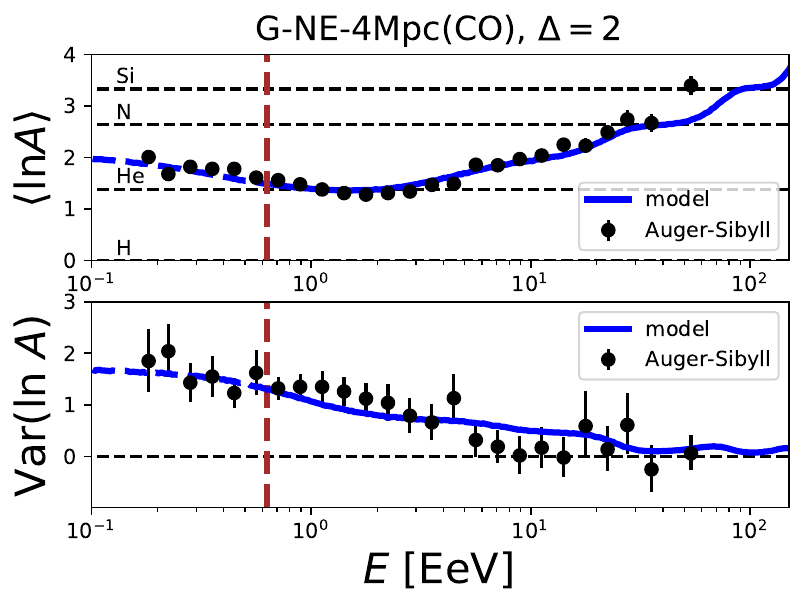}\\
    \includegraphics[width=0.4\textwidth]{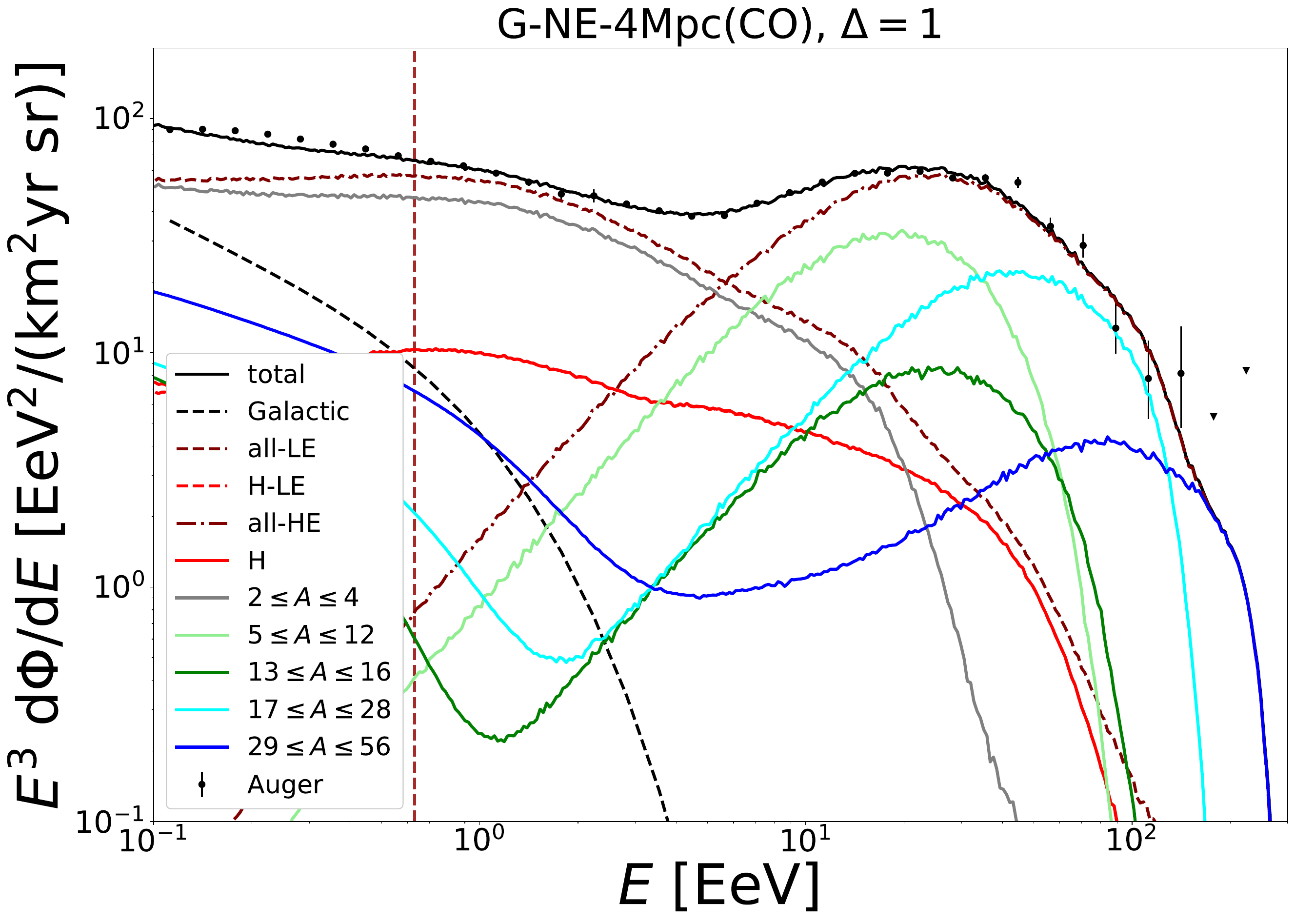}\includegraphics[width=0.4\textwidth]{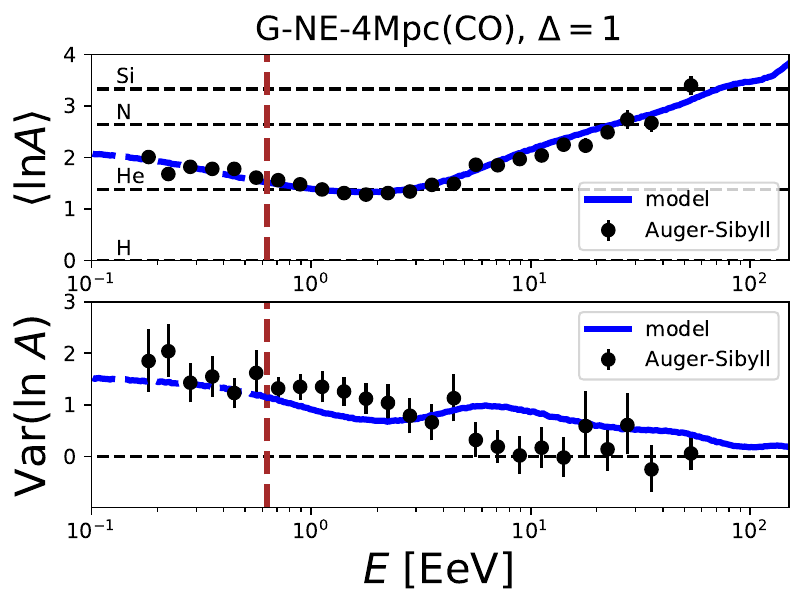}\\
   
    \caption{As in Fig.\,\ref{fD2noG} but considering C and O rather than N for the HE population, both for $\Delta=2$ (top) and 1 (bottom).}
    \label{fGCO}
\end{figure}

Let us note that for a continuous  distribution of sources, the averaging out of the spectral features resulting from the contributions from sources at different redshifts implies that there is not much difference between the results obtained assuming that the CNO component consists just of N nuclei or instead of C and O nuclei, and we will hence not report those results. On the other hand, for the case of a single nearby source the results can have noticeable differences.
We will then allow for six elements to get injected by the nearby source: H, He, C, O, Si and Fe, labelling this scenario as 4Mpc(CO), while for the LE sources (and the Galactic population) we still consider that the injected elements are H, He, N, Si and Fe.

We show  in Fig.~\ref{fGCO} and Table~\ref{tGCO} the results of the fit, now for both the cases with $\Delta=2$ and 1.

 For the case with $\Delta=2$ one can see that the  C nuclei dominate the suppression feature and the O nuclei help to reproduce the features of the spectrum at slightly higher energies. The quality of the fit further improves in this case, leading to  $\chi^2/{\rm dof}\simeq 4$.  The other qualitative features of this scenario are similar to those found previously.

We turn now  to discuss the case with a broader cutoff shape having  $\Delta=1$. In this case harder spectra and lower rigidity cutoff values are in general  inferred for the HE  component. Although the nominal value of the cutoff is smaller, given the harder shape of the source spectrum the different elements may anyhow reach  maximum energies comparable to those achieved in the case with $\Delta=2$, that had a larger nominal cutoff. One can also see that some qualitative differences appear with respect to the $\Delta=2$ case, since the fit prefers to have a large low-energy cutoff $R_{\rm cut}^{\rm L}\gg 10$~EeV and the presence of a broad C bump in the spectrum of the nearby source leaves little space for a He component, so that actually most of the He around the instep feature results from the LE population.  Some Si contributes near the suppression but anyway the features in the spectrum near the suppression are not very well reproduced with this broad cutoff shape.
Let us mention that  values of $\Delta$ smaller than 1 or bigger than 2  would not   significantly improve the quality of the fit.

\section{Including magnetic field effects and a finite source lifetime}

\begin{figure}[t]
    \centering
    \includegraphics[width=0.7\textwidth]{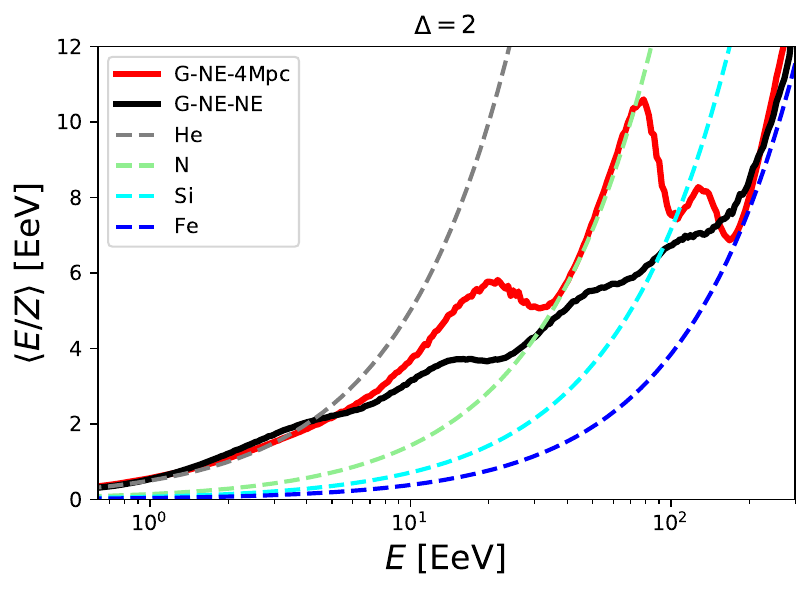}\\    
    \caption{ Average rigidity vs. energy for the models  in Fig.~\ref{fD2G}. Dotted lines indicate the rigidities corresponding to pure compositions for different elements.}
    \label{frigidityvse}
\end{figure}

In the scenarios discussed before in which no magnetic field effects were accounted for, the flux from the nearby source would be expected to be very anisotropic at ultrahigh energies, coming essentially from the direction of the source, which is at variance with observations. However, extragalactic and Galactic magnetic field induced deflections will change this picture. In order to get an insight into the typical CR rigidities present at the different energies, which ultimately determine the average magnetic deflections expected, we show in Fig.~\ref{frigidityvse} the values of $\langle E/Z\rangle$ as a function of energy for the scenarios  with  $\Delta=2$ displayed in  Fig.~\ref{fD2G}. One can appreciate that $\langle E/Z\rangle$ shows a mild growth with increasing energy and  it remains below  10~EeV in all  the  energy range observed. The nearby source scenario shows anyhow a more pronounced bump near few tens of EeV, due to the fact that in this case there are less photodisintegrations en route than in the NE-NE scenario, and hence at those energies the N component is more pure. These results are qualitatively similar also in the other variations of these scenarios that we consider.

A turbulent EGMF is expected to lead to a blurring of the source image,
with a characteristic radial angular spread of \cite{ha02,ha16}
\begin{equation}
    \theta_{\rm rms} \simeq 14^\circ \frac{10\,{\rm EeV}}{E/Z}\frac{B_{\rm rms}}{10\,{\rm nG}} \sqrt{\frac{l_{\rm coh}}{100\,{\rm kpc}}\frac{r_{\rm s}}{4\,{\rm Mpc}}} ,
    \label{eqthetarms}
\end{equation}
with $\theta$ being the angle between the CR arrival  direction and the line of sight to the source and where  the coherence length is typically expected to be within the range 10\,${\rm kpc}<l_{\rm coh}<1$\,Mpc.
Hence, if the extragalactic turbulent component within 4~Mpc has a root mean square strength satisfying $B_{\rm rms}\sqrt{l_{\rm coh}}\simeq 20\,{\rm nG}\sqrt{100\,{\rm kpc}}$, the angular spread should be of order 30$^{\circ}$  at the maximum achievable CR rigidities, corresponding to $E/Z\simeq 10$\,EeV, and it will increase for lower rigidities. This can explain why no anisotropy signals have been detected on  angular scales of few degrees, with just some hints of anisotropies having been  obtained for intermediate angular scales of about 30$^\circ$  at energies larger than 40~EeV. For decreasing energies the particles should progressively tend to a diffusive regime, and an accurate expression for the average deflections with respect to the source direction, in the case of a steady source,  is given by \cite{ha16}
\begin{equation}
    \langle \cos\theta\rangle = \frac{1-\exp(-3R-3.5R^2)}{3R}\equiv C(R),
\end{equation}
where $R\equiv r_{\rm s}/l_D$, with the diffusion length being well reproduced, in the case of a Kolmogorov turbulence spectrum, by the expression
\begin{equation}
    l_D\simeq l_{\rm coh}\left[4(E/E_{\rm c})^2+0.9(E/E_{\rm c})+0.23  (E/E_{\rm c})^{1/3}\right],
\end{equation}
with $E_{\rm c}\simeq 0.9Z(B_{\rm rms}/{\rm nG})(l_{\rm coh}/{\rm Mpc})$\,EeV being the critical energy for which the Larmor radius equals the coherence length, so that the regime of resonant diffusion takes place for $E<E_{\rm c}$.

Note that in order to account for the total CR density observed above 10~EeV, which is about $3\times 10^{-13}$\,km$^{-3}$,  an individual steady source at a distance  $r_{\rm s}=4$~Mpc and in the absence of magnetic fields would need to have an associated luminosity of ${\cal L}(E>10\,{\rm EeV})\simeq 5\times 10^{41}$\,erg\,s$^{-1}$, which is well within the typical luminosities of the AGN jets of $10^{43}$ to $10^{45}$\,erg\,s$^{-1}$ (with the Cen~A jet having  at present a power close to the lower value).
On the other hand,  for a steady isotropically emitting source whose CRs get dispersed by turbulent magnetic fields, the CR density at a given distance is actually enhanced with respect to the expectations from rectilinear propagation by an amount $\xi=\langle\cos\theta\rangle^{-1}$ \cite{mo19}. 
This local density enhancement would hence proportionally reduce the requirement on the CR source luminosity, typically by one order of magnitude. In addition, one should keep in  mind that a fraction of the events above 10\,EeV  will actually  be due to the LE population.

We  assumed up to now  that the source was steady, but in a more realistic situation the source activity will be time dependent. Note that if the acceleration were  a transient effect involving short times, one would anyhow expect a dispersion in the arrival times  resulting from the propagation across the turbulent fields encountered. In particular, the CRs from a bursting source emitted at one given time will arrive  to the Earth spread over a much longer time, which for not too large deflections is of order $\delta\tau\simeq r_{\rm s}\theta_{\rm rms}^2/6c\simeq 2\theta_{\rm rms}^2(r_{\rm s}/4\,{\rm Mpc}) 10^6$\,yr, with the angle expressed in radians.\footnote{This just considers the difference between the straight propagation and that along an arc of a circle arriving to the observer at an angle $\theta_{\rm rms}$ with respect to the source direction.}
The energy dependent time dispersion effects  will also affect the spectral shape (see e.g. \cite{mi96}), in a way which would depend on the detailed emission history of the source. Note that this time spread could also lead to an overlap of different emission pulses if their temporal separation happens to be smaller than the propagation induced time spread. 

We will consider for definiteness  that the source has been emitting steadily since an initial time $t_i$. This could be for instance an approximation to the scenario in which the acceleration is related to the activity of the AGN jets, which typically have expected lifetimes of few hundred million years and may have started after a collision of two galaxies or after an enhanced episode of accretion.  In this case, analytic expressions have been obtained for the energy dependent flux enhancement factor  that is induced by the extragalactic magnetic field \cite{ha21}.
The main effect of the finite source lifetime will be to suppress the diffusive enhancement of the flux  at low rigidities, when the propagation time of the CRs becomes larger than the source lifetime. An accurate fit to the enhancement factor is in this case
\begin{equation}
    \xi = \frac{1}{C(R)}\exp\left[-\left(\frac{r_s^2}{0.6 l_D \,ct_i}\right)^{0.8}\right].
\end{equation}

\begin{figure}[t]
    \centering
    \includegraphics[width=0.7\textwidth]{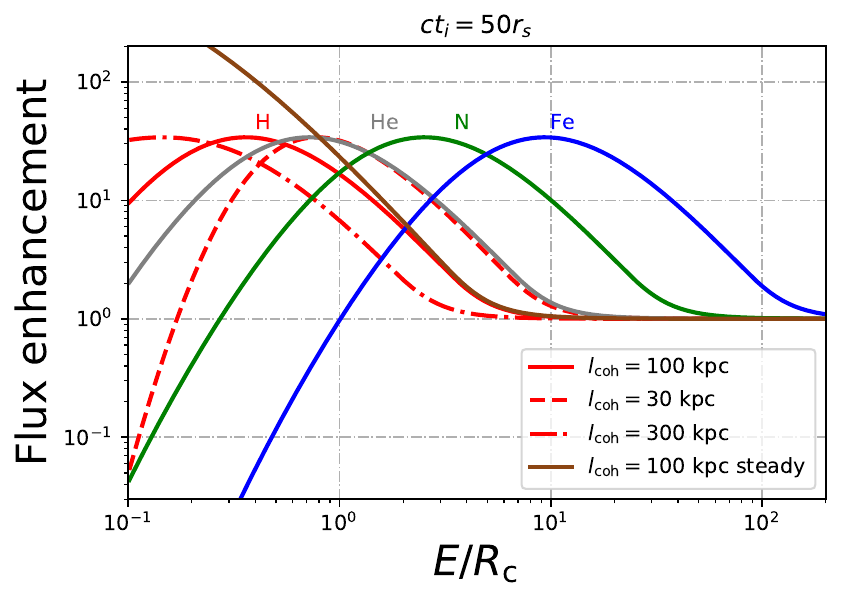}\\    
    \caption{ Enhancement factor as a function of $E/R_{\rm c}$ considering that $ct_i=50r_{\rm s}$. Results are shown for different nuclei with $l_{\rm coh}=100$~kpc and  for different values of $l_{\rm coh}$ in the case of H nuclei. }
    \label{fenhancement}
\end{figure}
In Fig.~\ref{fenhancement} we show the flux enhancement factor that is expected for CRs diffusing from a source at 4~Mpc distance which is emitting since $t_i =50r_s/c$, for H nuclei with different values of $l_{\rm coh}$. The results are also shown for different nuclei for the case with $l_{\rm coh}=100$\,kpc, whose enhancement appear at energies scaled simply by the corresponding charges, and are plotted as a function of $E/R_{\rm c }$, where we introduced  $R_{\rm c}\equiv E_{\rm c}/Z\simeq 0.9 (B_{\rm rms}/{\rm nG})(l_{\rm coh}/{\rm Mpc})$\,EeV. For reference, also the enhancement for a steady source for the case of H nuclei  is shown with a brown line, in which case no suppression appears at low energies. 

Let us mention that one important difference between the single source model and the one based on a distribution of sources up to high redshifts is that in the later case no  enhancement of the flux due to the diffusion in the EGMF will result, since when the nearby sources get enhanced the faraway ones get suppressed, and hence  no significant overall amplification results (this is just the propagation theorem \cite{al04}). Only at low energies, when all the sources get suppressed,  the magnetic horizon suppression will show up if we account for the fact  that the closest source is at a finite distance.

Note that in the high-energy side, the flux enhancement shown in Fig.~\ref{fenhancement} starts to become significant when the propagation ceases to be quasi-rectilinear, i.e. for $l_D<r_{\rm s}$, which corresponds to $E<E_{\rm rect}=0.5 \sqrt{r_{\rm s}/l_{\rm coh}}E_{\rm c}$. For the cosmic rays with energy $E_{\rm rect}$, the average CR deflection along their trajectory would be $\delta \simeq 1$~rad, and hence the resulting angle at Earth with respect to the line of sight to the source will be on average $\theta\simeq \delta/\sqrt{3}\simeq 33^\circ$. In order to match the angular size of the observed hot spot  we would like that deflections of this size appear at rigidities close to $R_{\rm rect}\equiv E_{\rm rect}/Z\simeq 10$~EeV (i.e. for N nuclei of 70~EeV). On the other hand, one can obtain the following expression for the critical rigidity  
\begin{equation}
  R_{\rm c} \simeq 3.2\,{\rm EeV}\frac{R_{\rm rect}}{10~{\rm EeV}} \sqrt{\frac{l_{\rm coh}}{\rm 100\,kpc}\frac{4\,{\rm Mpc}}{r_{\rm s}}}.
\end{equation}

For the source with a finite lifetime, the enhancement reaches a maximum value $\xi_{\rm max}\simeq 0.8 ct_i/r_{\rm s}$ at an energy $E_{\rm max}\simeq 0.5r_{\rm s}/\sqrt{ct_il_{\rm coh}}E_{\rm c}$ \cite{mo19}. Since the dipolar modulation of the flux of a given mass component is directly related to the enhancement factor, with the dipolar amplitude at the energy of the maximum enhancement being $d(E_{\rm max})\simeq 2.7/\xi_{\rm max}\simeq 3.4r_{\rm s}/ct_i$, in order not to exceed the dipolar amplitudes measured above 8~EeV one needs that a significant density enhancement should be achieved. For instance, requiring that $d(E_{\rm max})< 0.1$ this would require $ct_i> 30r_{\rm s}$, and we will hence adopt the reference value $ct_i= 50r_{\rm s}$ as a guideline. Note that this time is also comparable to the age of the Cen~A inner lobes of about $6\times 10^8$~yr \cite{is98}. Combining the above expressions one can then see that the rigidity at which the maximum enhancement is achieved is just 
\begin{equation}
    R_{\rm max}\simeq 1.4\,\frac{R_{\rm rect}}{10\,{\rm EeV}}\sqrt{\frac{50r_{\rm s}}{ct_i}}\,{\rm EeV},
\end{equation}
and then for CNO nuclei this rigidity would  correspond to energies just above the ankle one. On the other hand, the required EGMF strength in this scenario can be expressed as 
\begin{equation}
    B_{\rm rms}\simeq 33 \frac{R_{\rm c}}{3\,{\rm EeV}}\frac{100\,{\rm kpc}}{l_{\rm coh}}\,{\rm nG}.
    \label{eqbrms}
\end{equation}

Another important effect impacting on the anisotropies of the CR arrival directions is related to the deflections by the Galactic magnetic field, which has a global coherent structure with a disk component following the spiral structure as well as some halo components, and in addition a turbulent component should also be present. In particular, the regular Galactic magnetic field will give rise to lensing effects \cite{ha99}, which can magnify or demagnify the source image as well as deflect it and, more remarkably, they give rise to multiple images of the source, with the new images appearing in pairs in a  direction of the sky different  from that of the primary deflected source image. The random Galactic magnetic field component will also further blur the different images. However,  in the scenarios we consider the effect of the Galactic turbulent field will be subdominant with respect to that of the turbulent EGMF, which although having a smaller amplitude it acts on much larger distances and has a larger coherence length, and we will hence neglect the turbulent Galactic magnetic field component. 

When the regular Galactic magnetic  field effects are accounted for, the source will be seen with a very different appearance from  that of a single blurred image. One can indeed expect to see many different blurred and distorted images appearing in different regions of the sky, some possibly very far away from the principal image that is originally present before the Galactic deflections enter into play. This scenario, including both the EGMF and the Galactic one,  was discussed in detail in \cite{mo22}, showing that one can obtain a CR arrival direction distribution in line with the observations, having essentially a dipolar distribution at the level of 10\% at 10\,EeV and then  some hot spots at intermediate angular scales at the highest energies.\footnote{Reference \cite{mo22} considered besides Cen~A also the possibility of having a smaller contribution from the source M82, which lies at a similar distance, so as to account for a localised excess near the direction of that source hinted by the Telescope Array observatory \cite{ab18,fu21}. However, data from the Auger Observatory does not show similar flux excesses in those directions \cite{go23b}, and we hence include here just the contribution from Cen~A.  See also ref.~\cite{pf17} for a scenario to account for these two localised anisotropies using Cen~A and M82.} 

Let us mention that for realistic Galactic magnetic field models, multiple images of some extragalactic sources can already appear at $E/Z\simeq 30$\,EeV, while for $E/Z\simeq 10$\,EeV most of the sky directions will lead to multiple images and for  $E/Z\leq 5$~EeV already several images  are expected for any given source \cite{ha99} (see also \cite{ke15} for a study of the deflections for the specific direction of Cen~A). In particular,  for the nearby source this implies that at energies of  tens of EeV the dominant CNO component should be in the regime of several images, with the heavier elements being even more isotropised. On the other hand, the He
component may be instead near the regime of one or few images as one approaches the highest energies that it can reach, of about 20~EeV, and hence its possible impact on the anisotropies  will deserve further consideration (see next section).

\begin{figure}[t]
    \centering
     \includegraphics[width=0.4\textwidth]{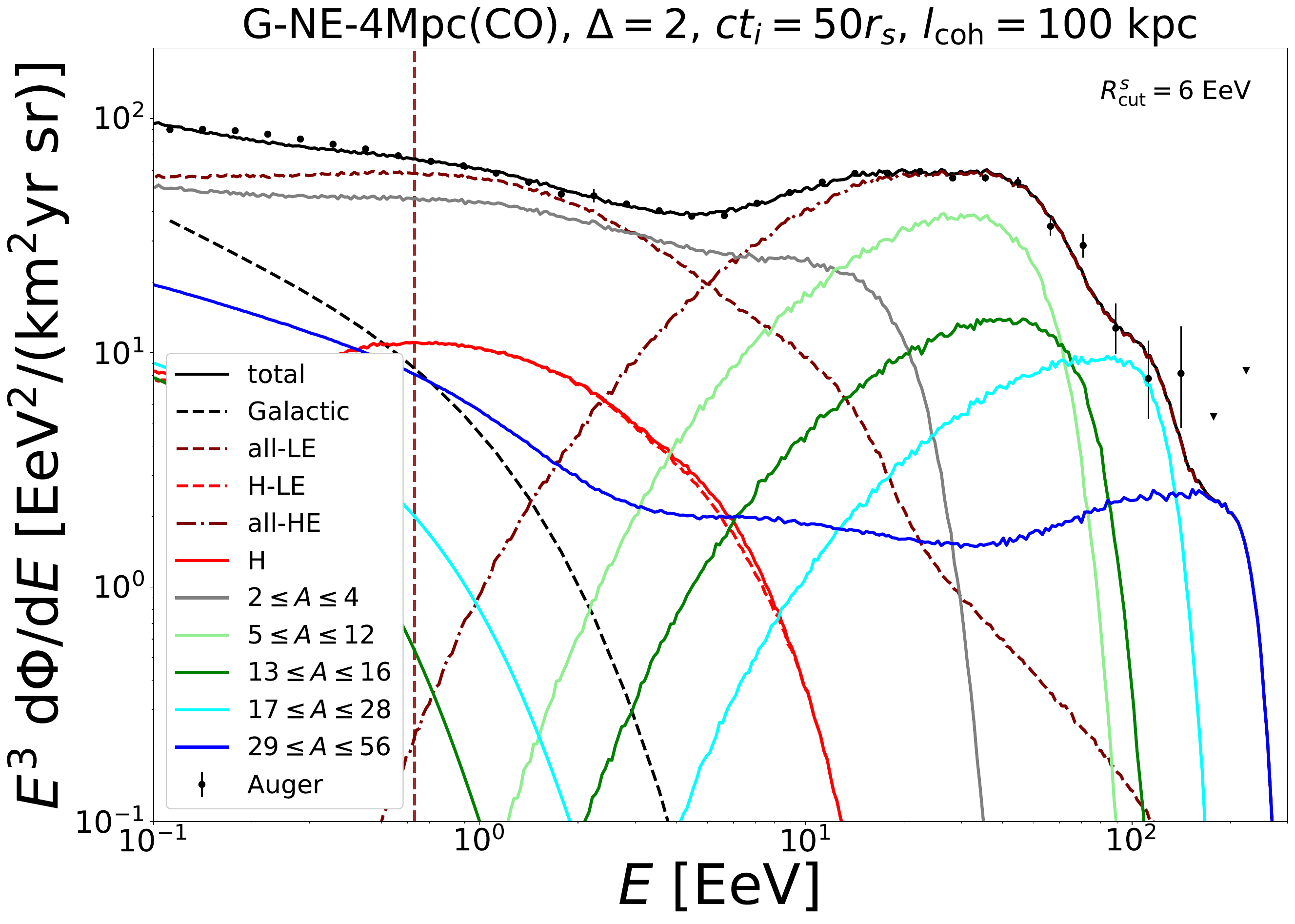}\includegraphics[width=0.4\textwidth]{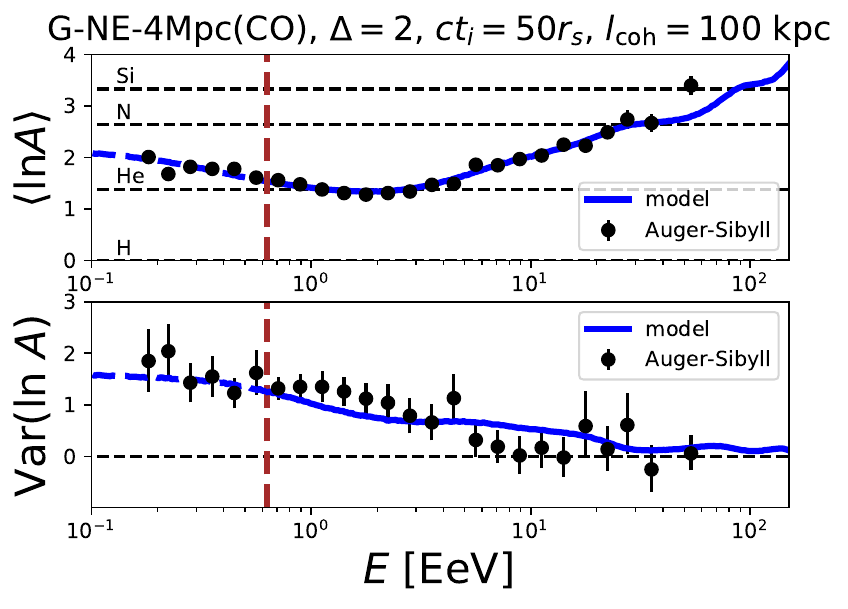}\\

      \includegraphics[width=0.4\textwidth]{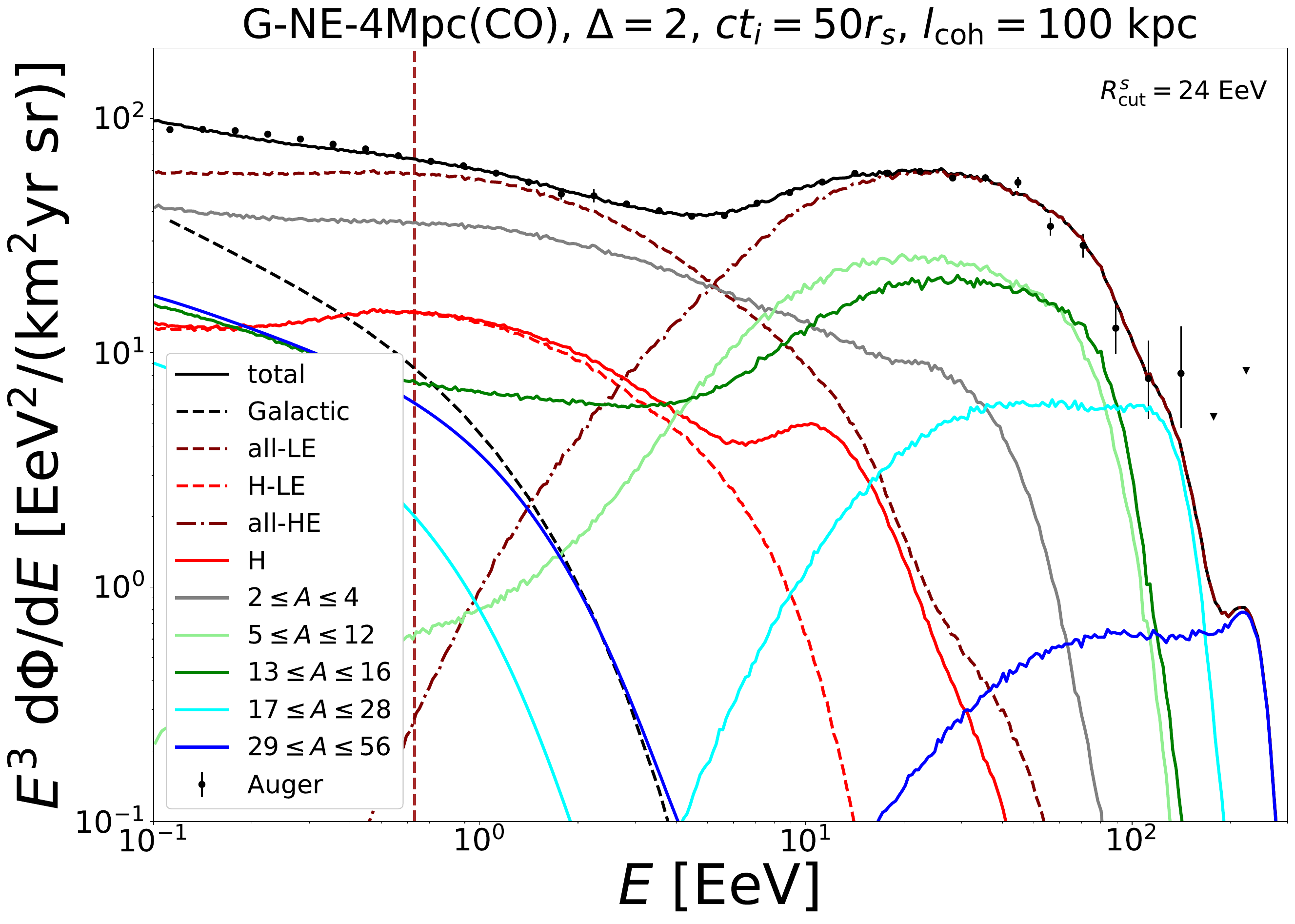}\includegraphics[width=0.4\textwidth]{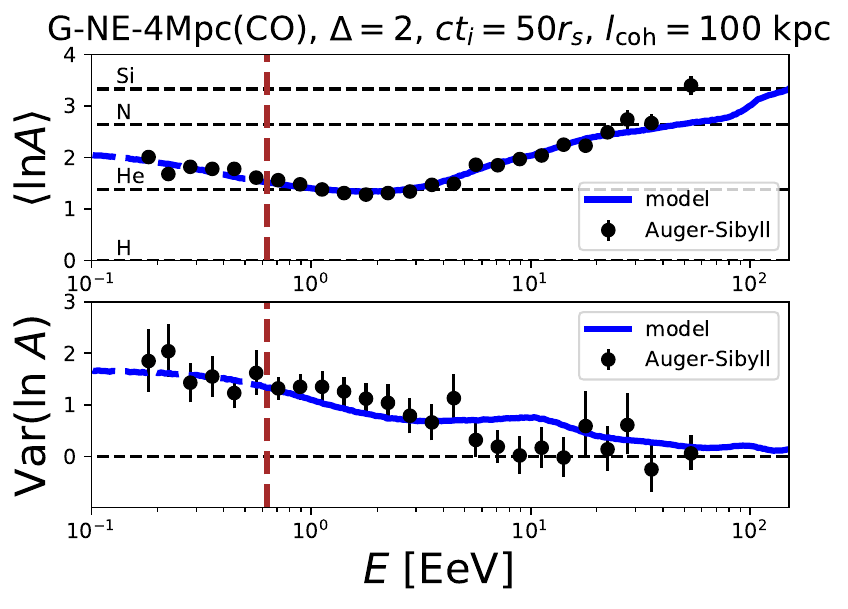}\\
    
      \includegraphics[width=0.4\textwidth]{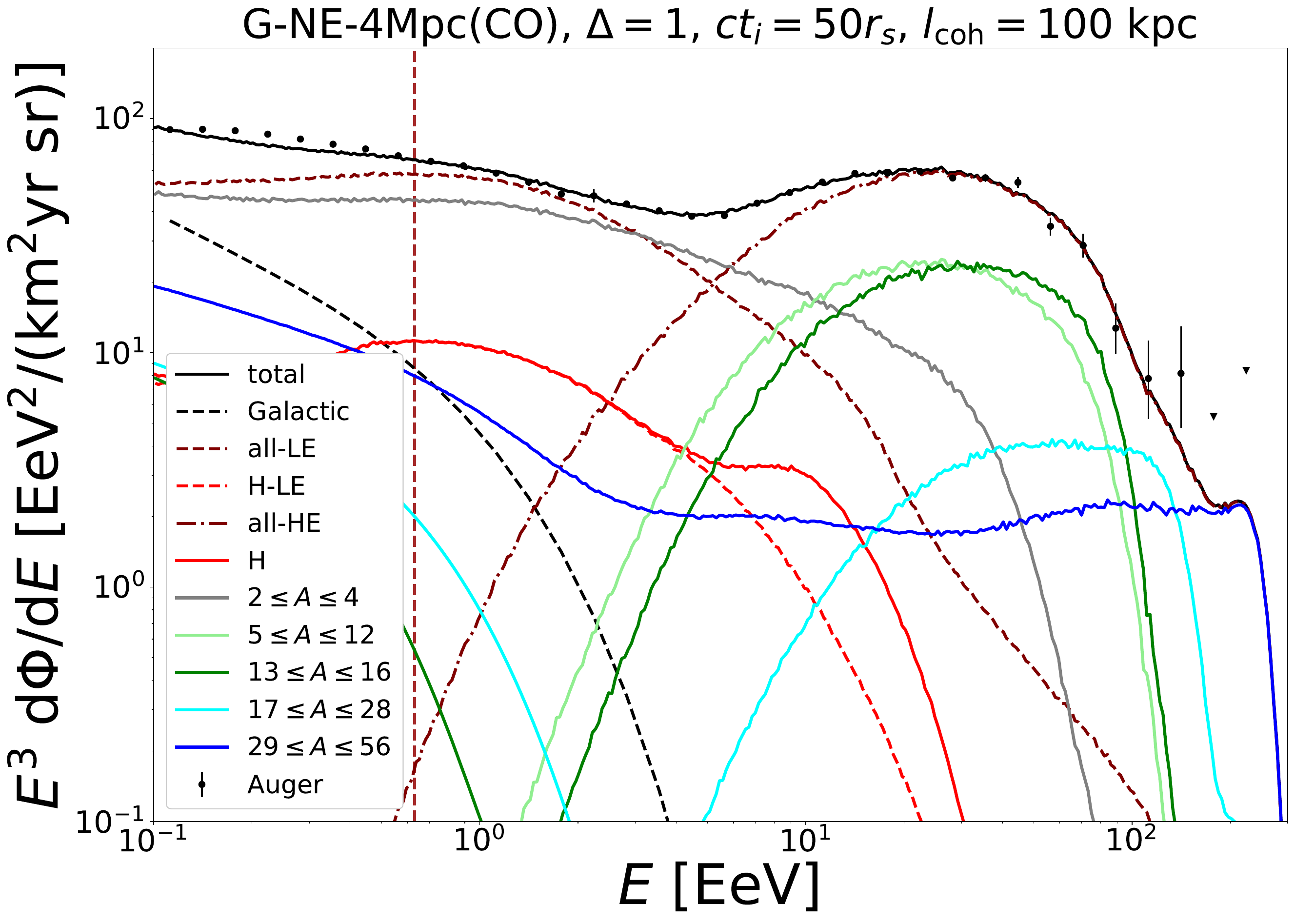}\includegraphics[width=0.4\textwidth]{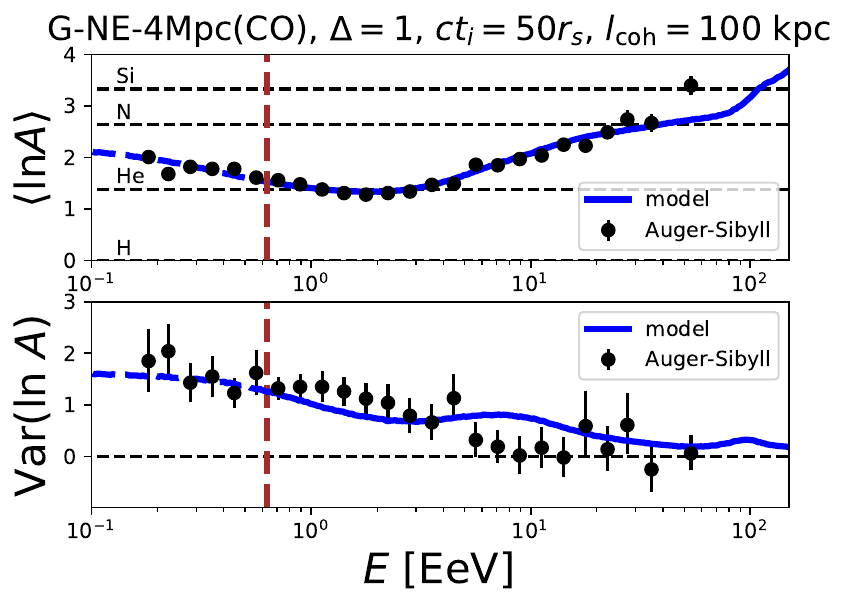}\\
  
    \caption{Spectrum and composition in the scenarios including the effects of EGMF  with $l_{\rm coh}=100$~kpc and a source emitting steadily since an initial time $t_i=50r_{\rm s}/c$. Two different results are shown for the case with $\Delta=2$ (see text).}
    \label{fGCOB}
\end{figure}

The magnification effects of the Galactic magnetic field could in principle also affect the observed spectrum, eventually producing features in it. However, given that the source image is already significantly blurred by the EGMF, one also expects that the magnification peaks associated to the appearance of new images, which in the limit of a pointlike source are actually divergent \cite{ha00}, will be to a large extent  smoothed out. In particular, in the limit in which the  CR flux arriving to the Galaxy is isotropic, the Liouville theorem implies that the deflections by the regular Galactic magnetic field will not affect it. As a consequence,  the overall modifications of the spectral shape due to the Galactic magnetic field effects should not be strong in the regime with many multiple images, and we will hence only include the spectral distortions due to the diffusion in the EGMF shown in Fig.~\ref{fenhancement}, which should be the most important ones in the scenario considered.

In Fig.~\ref{fGCOB} and Table~\ref{tGCOB} we show the results of fitting the spectrum and composition with a model with a source at $r_{\rm s}=4$~Mpc with CO injection and emitting steadily since $t_i = 50r_{\rm s}/c$, considering a magnetic field with coherence length $l_{\rm coh}=100$~kpc and obtaining  from the fit also the value of $R_{\rm c}$ (and hence of the EGMF strength $B_{\rm rms}$ through Eq.~(\ref{eqbrms})). The results are computed both for  $\Delta=2$ and  $\Delta=1$. One can see that a good overall agreement with observations is obtained, even if the modulation of the flux due to the diffusion process is quite strong. The critical rigidity is found to be $R_{\rm c}\simeq 1.4$ to 4.8~EeV and hence the  required magnetic field strength in these cases is inferred to be  $B_{\rm rms}\simeq 15$ to 50\,nG. 
The $\chi^2/{\rm dof}$ is now about 3, which is  smaller than what obtained in the absence of magnetic field effects in the previous section. Note that for the case with $\Delta=2$ we report two qualitatively different solutions, which have comparable values of $\chi^2$. The first involves a value $R_{\rm cut}^{\rm s}=6$~EeV, and hence the rigidity cutoff  is the main responsible for the spectral suppression of the different source elements. The second one involves a much larger cutoff value, $R_{\rm cut}^{\rm s}=24$~EeV, and hence here it is the interactions with the CMB which mostly shape the spectral suppression. 
The fact that a solution with a large source cutoff appears when considering the magnetic field effects is made possible because in this case the  light secondaries (H and He), eventually produced by the photodisintegration of the C/O nuclei with the highest rigidities, tend to be buried under the diffusively enhanced flux of the C/O primaries of lower rigidities which are present at the same energy. Hence, the appearance of secondaries does not constrain anymore in a strong way the value of the rigidity cutoff of the source, which can then become large (of about 20 to 30~EeV). One can also appreciate that at the highest energies the Si contribution plays a relevant role, and one should keep in mind that its cutoff due to interactions starts close to 130~EeV.

\begin{table}[t]
\addtolength{\tabcolsep}{-1pt}
\centering

{\small
\begin{tabular}[H]{ @{}c| c  c c c c c c c| c c c c c|c| c @{}}
\multicolumn{16}{c}{ model G-NE-4Mpc(CO), $ct_i = 50 r_{\rm s}$, EGMF with $l_{\rm coh}=100$~kpc }\\
\hline
$\Delta$ & $\gamma_{\rm s}$ & $R_{\rm cut}^{\rm s}$ & $f^{\rm s}_{\rm H}$ & $f^{\rm s}_{\rm He}$ & $f^{\rm s}_{\rm C}$ & $f^{\rm s}_{\rm O}$ & $f^{\rm s}_{\rm Si}$ & $f^{\rm s}_{\rm Fe}$ &$\gamma_{\rm L}$ & $R_{\rm cut}^{\rm L}$ & $f^{\rm L}_{\rm H}$ & $f^{\rm L}_{\rm He}$ & $f^{\rm L}_{\rm N}$ & $R_{\rm c}$ &  $\frac{\chi^2}{\rm dof}$  \bigstrut[t]\\ 
 &  & [EeV] &  & & [\%] &  &  &  & &  [EeV]  &  & [\%] &  &  [EeV] &\bigstrut[b]\\
\hline
2& 0.4 & 6.0 & 0 & 86 & 12 & 2.1 & 0.3 & 0 & 3.0 & 8.2 & 13 & 83 & 0 & 1.4 & 2.90 \bigstrut[t]\\
 & 1.7 & 24 & 9.5 & 39 & 31 & 18 & 2.6 & 0.1 & 3.0 & 8.4 & 21 & 65 & 14 & 3.6 &3.07 \\
1 & 1.2 & 7.9 & 0 & 65 & 21 & 13 & 0.8 & 0.1 & 2.9 & 7.6 & 13 & 83 & 0 & 2.8 & 3.05\bigstrut[t]\\
\hline
\multicolumn{15}{c}{ idem with $E\times 1.14$} \bigstrut[t]\\
\hline
2& 1.9 & 29 & 27 & 17 & 32 & 20 & 3.1 & 1.1 & 3.0 & 10 & 17 & 63 &20 & 4.7 & 1.98 \\
 & 1.2 & 8.8 & 0 & 66 & 25 & 6.9 & 1.6 & 0.1 & 2.9 & 10.2 & 5.8 & 90 & 0 & 3.8 & 2.38 \\
1& 2.0& 23 & 27 & 13 & 32 & 23 & 2.9 & 1.5 & 3.0 & 11 & 16 & 64 & 19 & 4.8 & 2.18\\
\hline

\end{tabular}}

\caption{Parameters of the fit to the flux and composition for the different scenarios including an EGMF with $l_{\rm coh}=100$~kpc and a source emitting steadily since an initial time $t_i=50r_{\rm s}/c$, also considering emission of C and O from the nearby source rather than N.  Results are shown for $\Delta=2$ and 1, and also when rescaling the observed energies by a factor 1.14. }
\label{tGCOB}
\end{table}

\begin{figure}[t]
    \centering
     \includegraphics[width=0.4\textwidth]{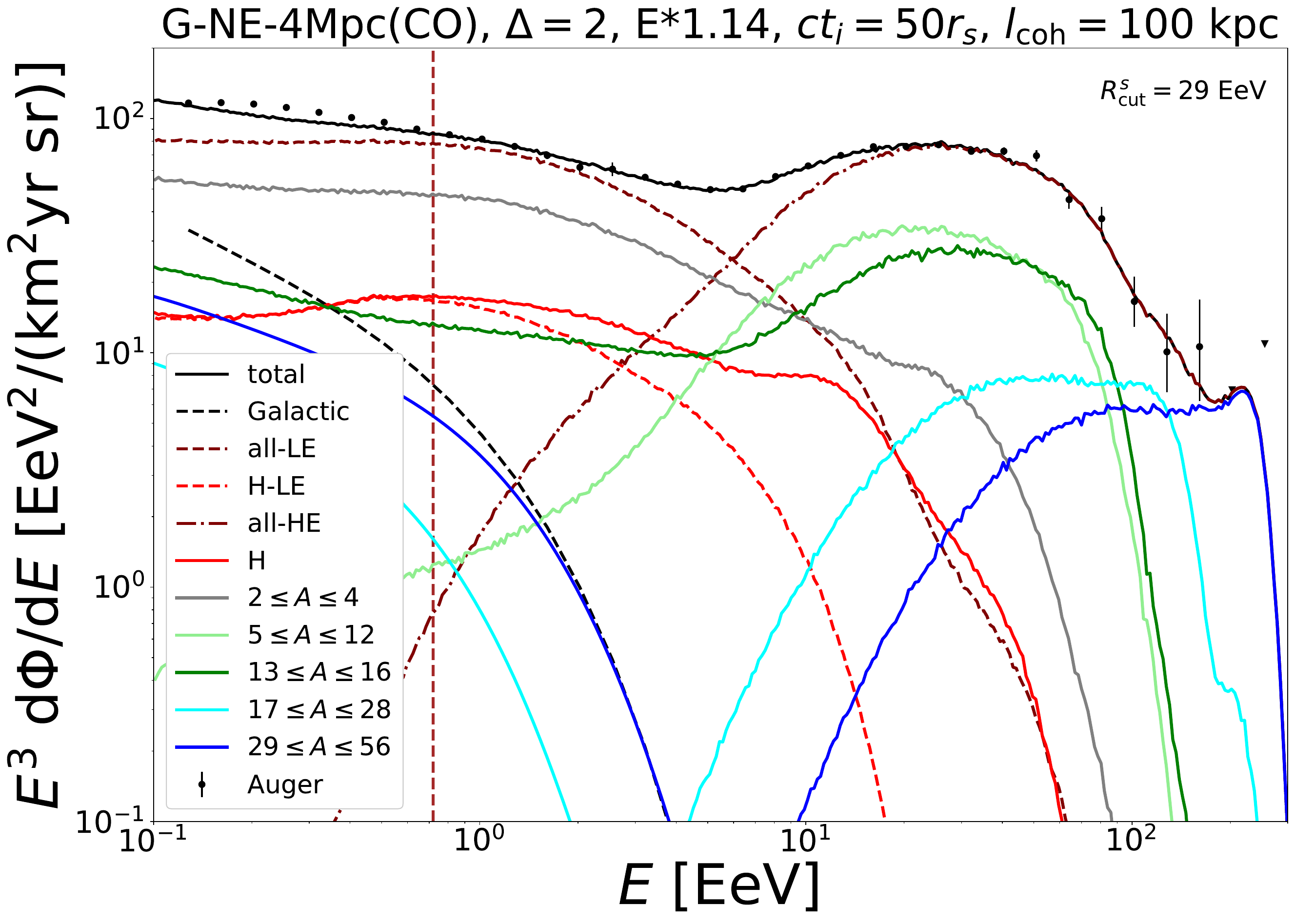}\includegraphics[width=0.4\textwidth]{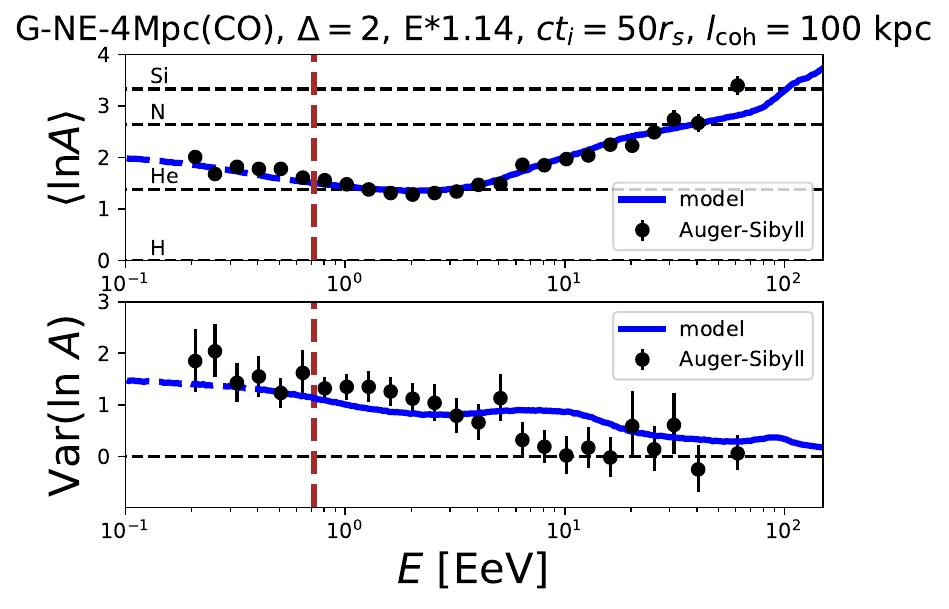}\\
     
      \includegraphics[width=0.4\textwidth]{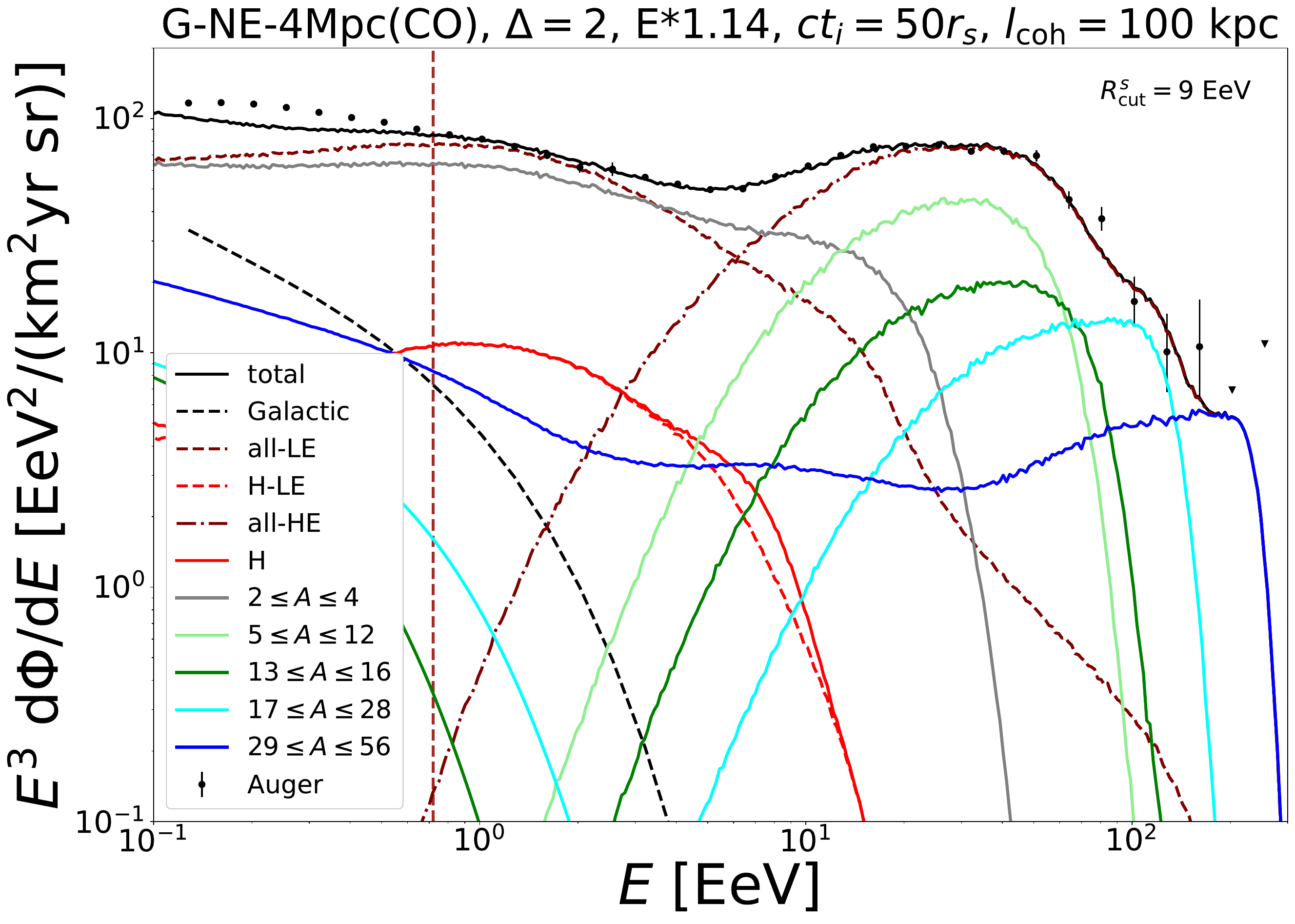}\includegraphics[width=0.4\textwidth]{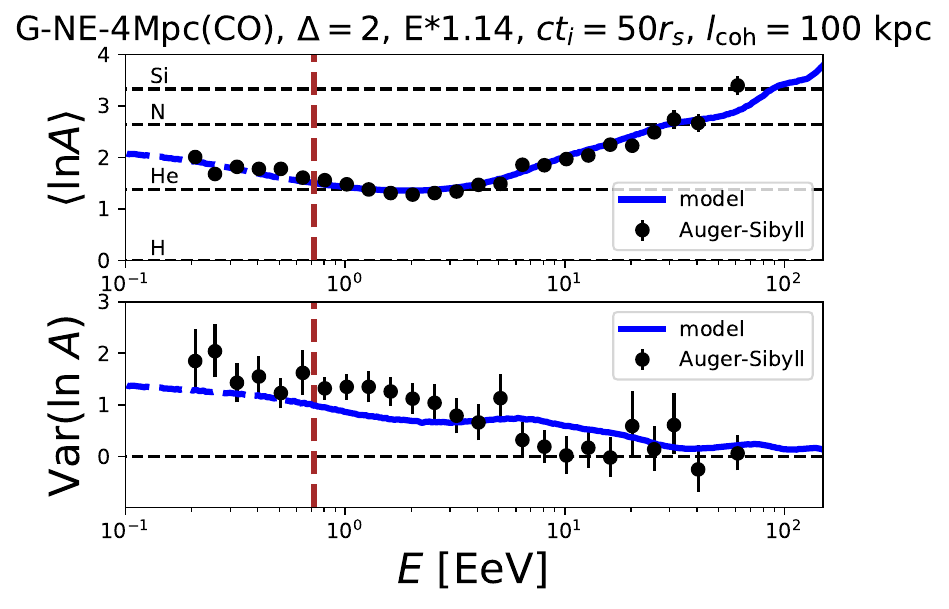}\\
    
       \includegraphics[width=0.4\textwidth]{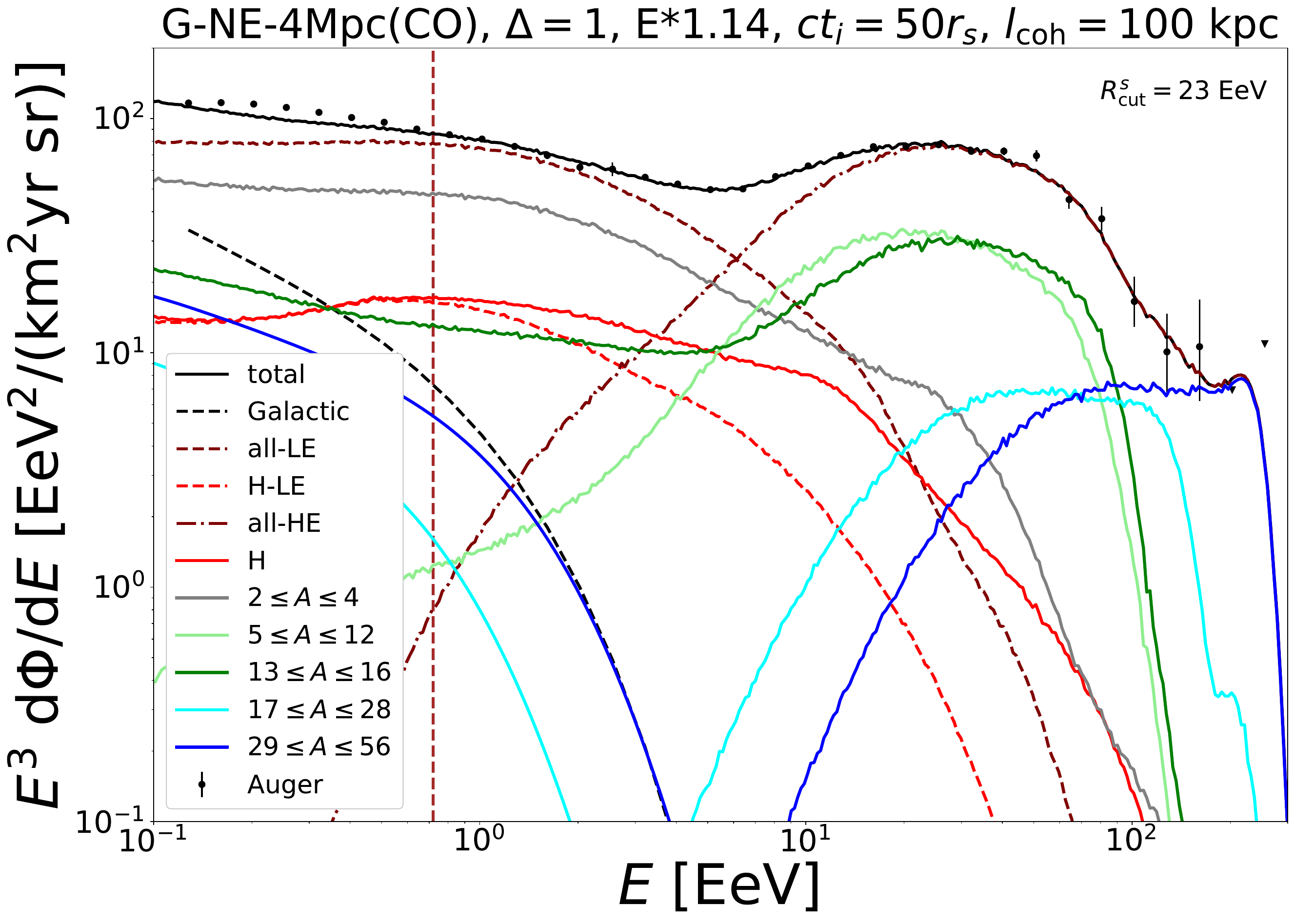}\includegraphics[width=0.4\textwidth]{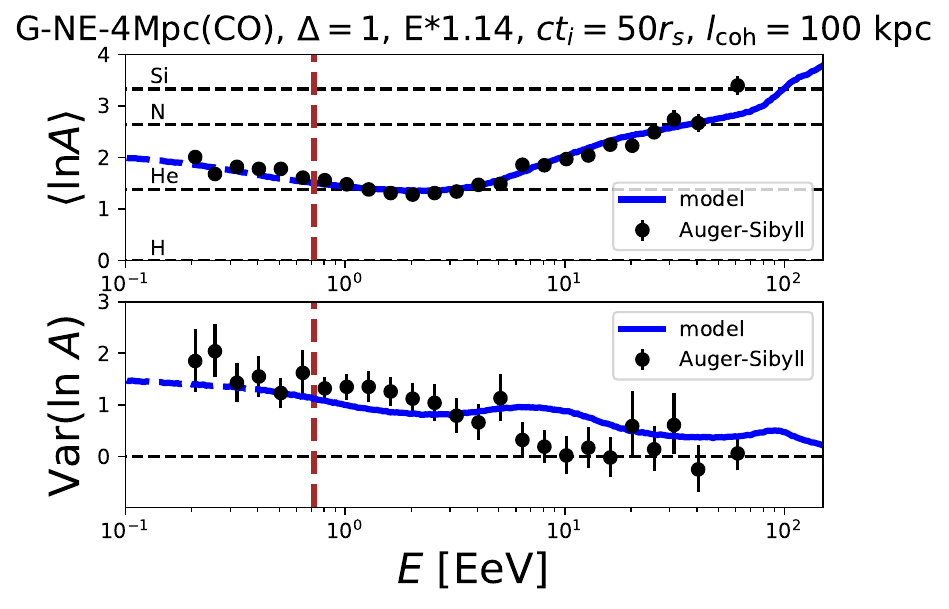}\\

    \caption{Same as Fig.~\ref{fGCOB} but with the measured energies multiplied by 1.14.}
    \label{fGCOEB}
\end{figure}

It is important to point out that systematic shifts in the energy calibration or in the $X_{\rm max}$ reconstructed shower depth would affect significantly the fits. In particular, the results obtained  considering
a systematic positive shift in the measured energies of $+14\%$, which corresponds to the quoted systematic uncertainty in the energy calibration of the experiment (and would  go in the direction of reducing the difference with the energy scale  of the Telescope Array experiment), are displayed in Fig.\,\ref{fGCOEB} and in the last two rows in Table~\ref{tGCOB}. One can see that  the $\chi^2$/dof  would improve in this case to about 2.0 to 2.4, which is much better.
In this case, the best fit scenario for both values of $\Delta$ would be the one with a large value of $R_{\rm cut}^{\rm s}$.  Since the high-energy suppression is here mostly due to the  effects  of the interactions with the CMB, the results depend very mildly on the actual shape of the cutoff adopted, with values of $\Delta=1$ or 2 leading to comparable values of the fitted parameters. One can also see that,  given the broad bump in the C/O components due to the diffusion, when shifting the ankle and instep features to higher energies less room is left for the He from the nearby source, whose fraction becomes of only about 10\%. Another particular feature is that the H component from the nearby source is no longer negligible, what is in part due to the reduced He contribution allowed in this case. Also the value of the spectral index of the nearby source turns out to be about 2, just around the value expected from diffusive shock acceleration.

\begin{figure}[t]
    \centering
     \includegraphics[width=0.49\textwidth]{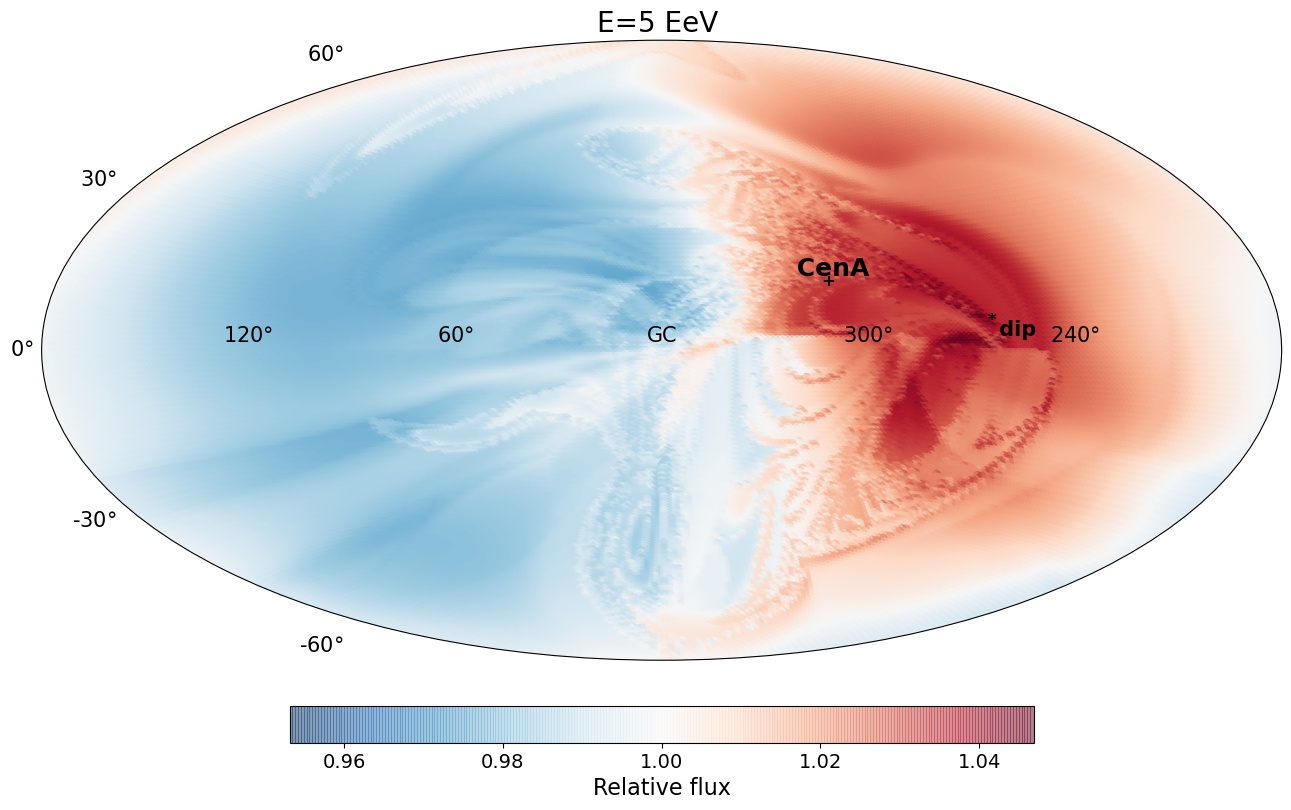}\includegraphics[width=0.49\textwidth]{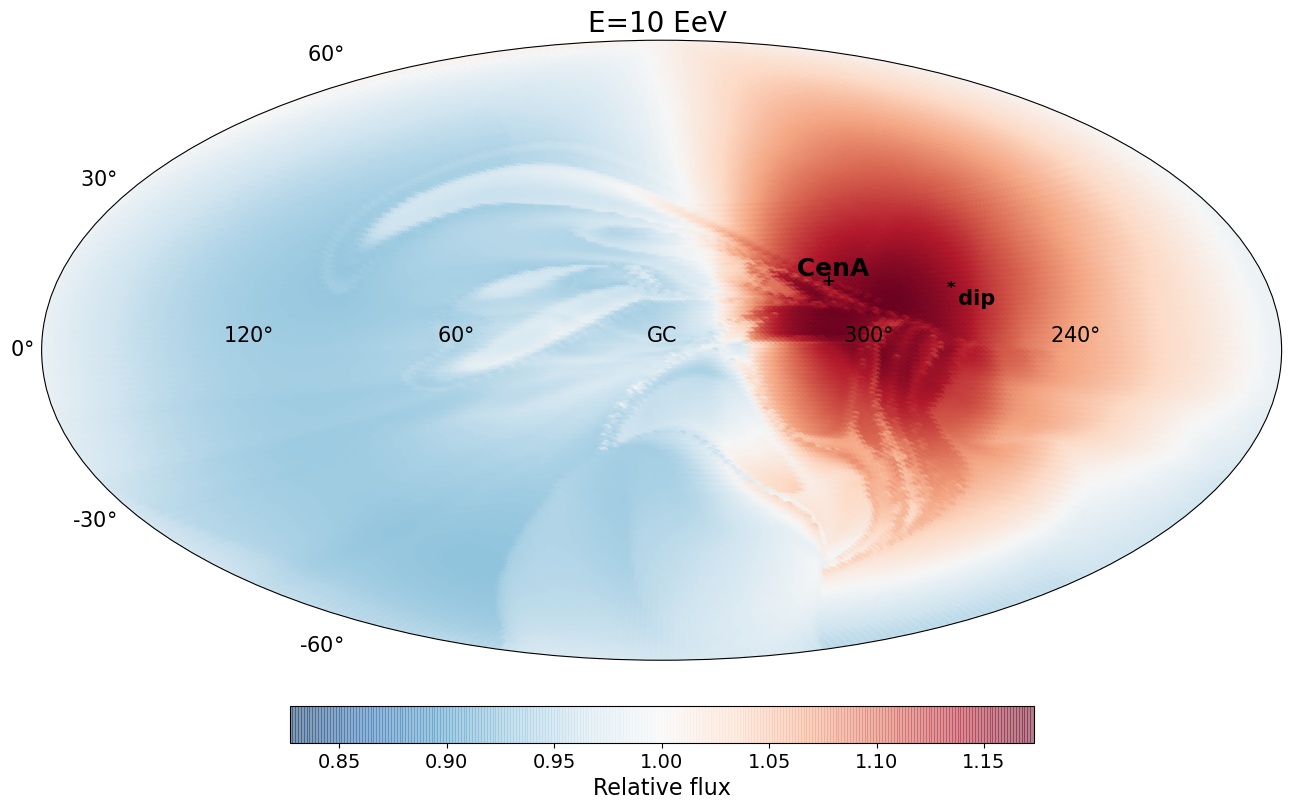}\\
  \includegraphics[width=0.49\textwidth]{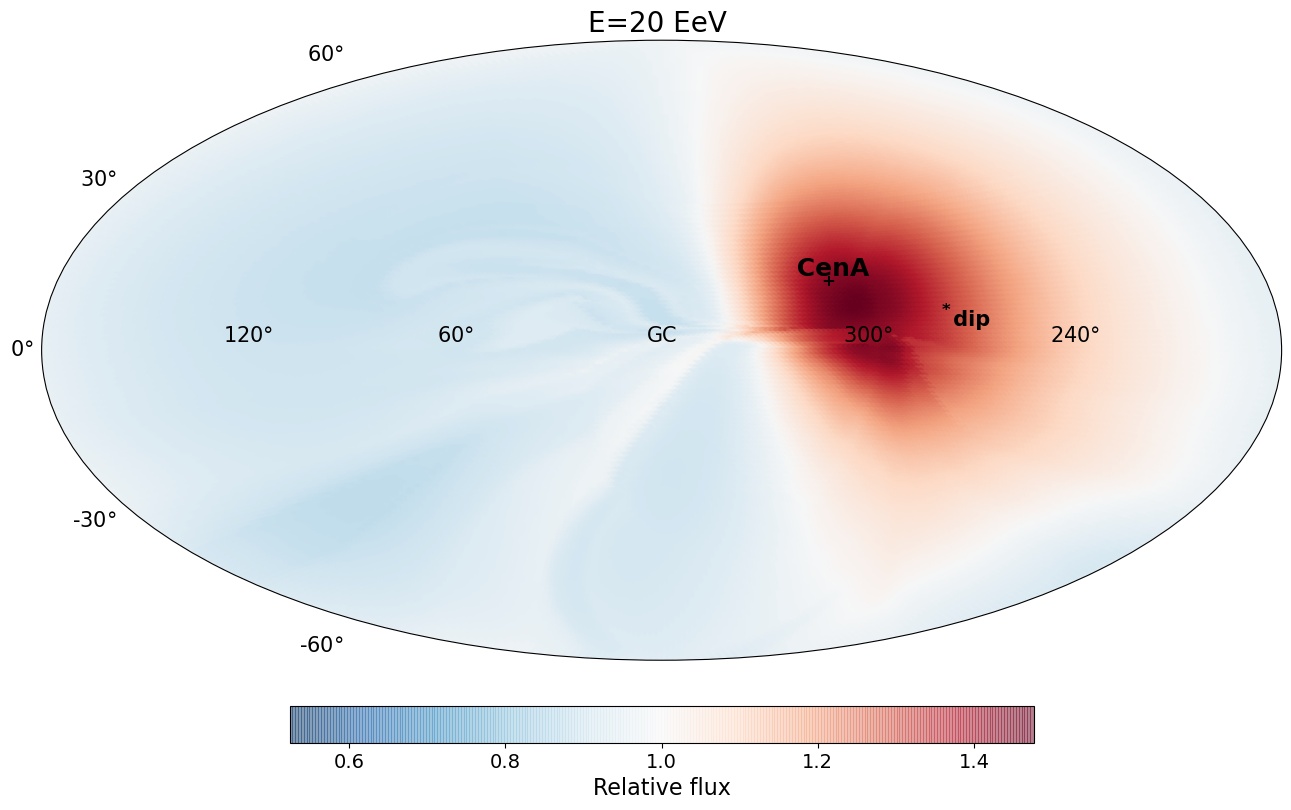}\includegraphics[width=0.49\textwidth]{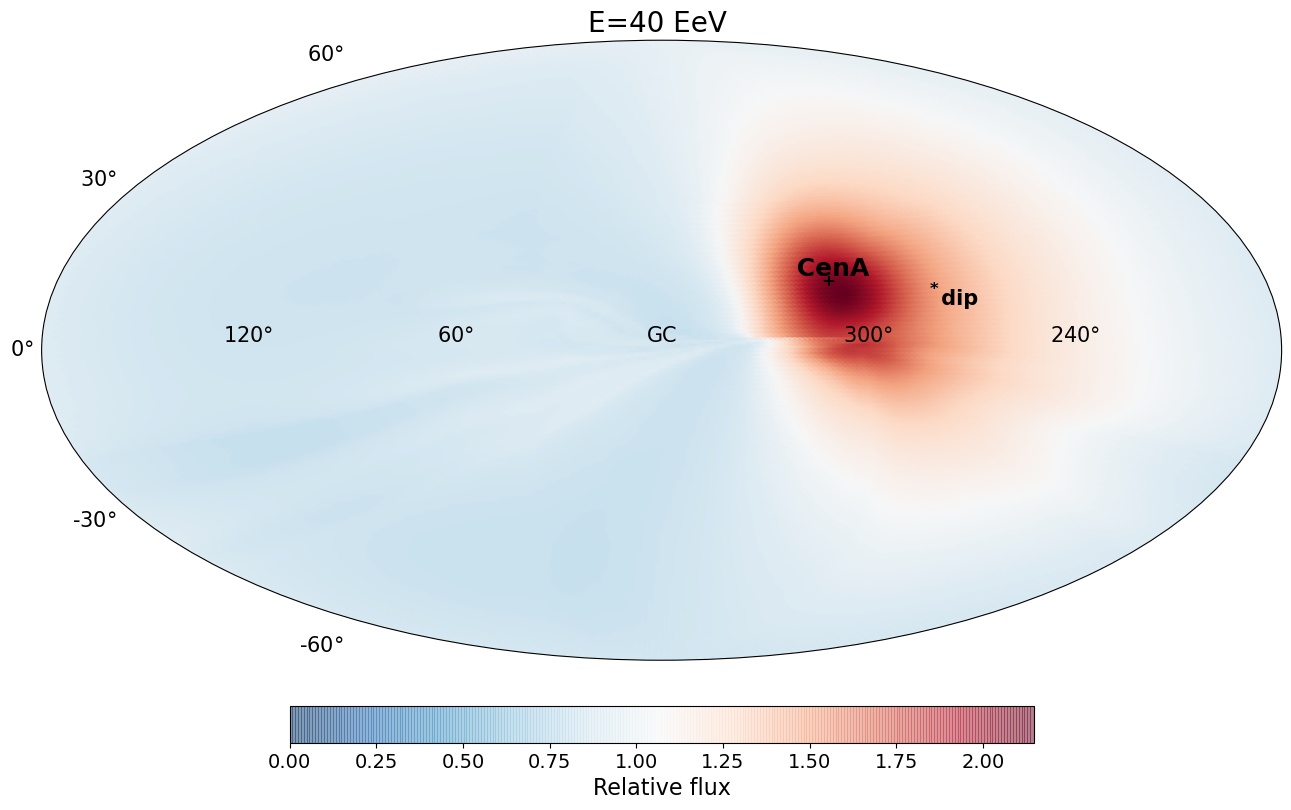}\\
  \includegraphics[width=0.49\textwidth]{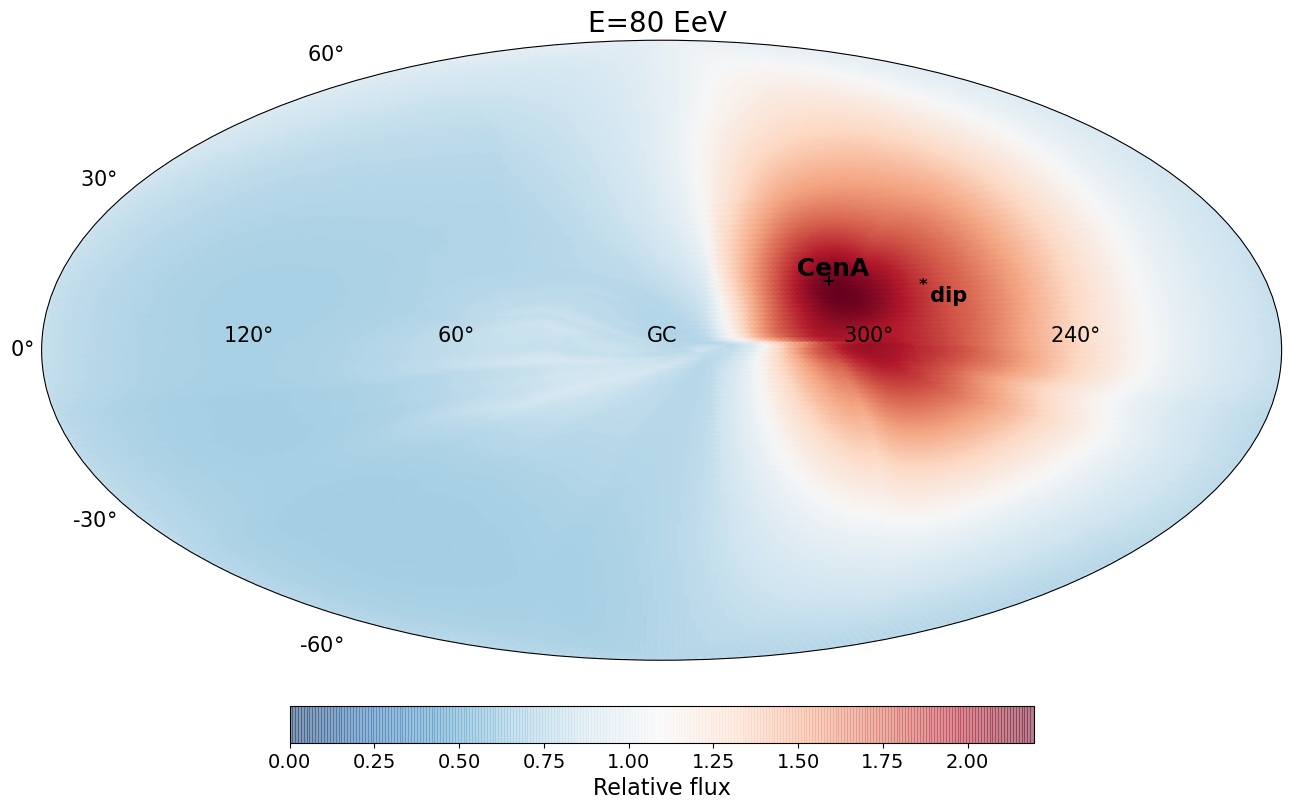}\includegraphics[width=0.49\textwidth]{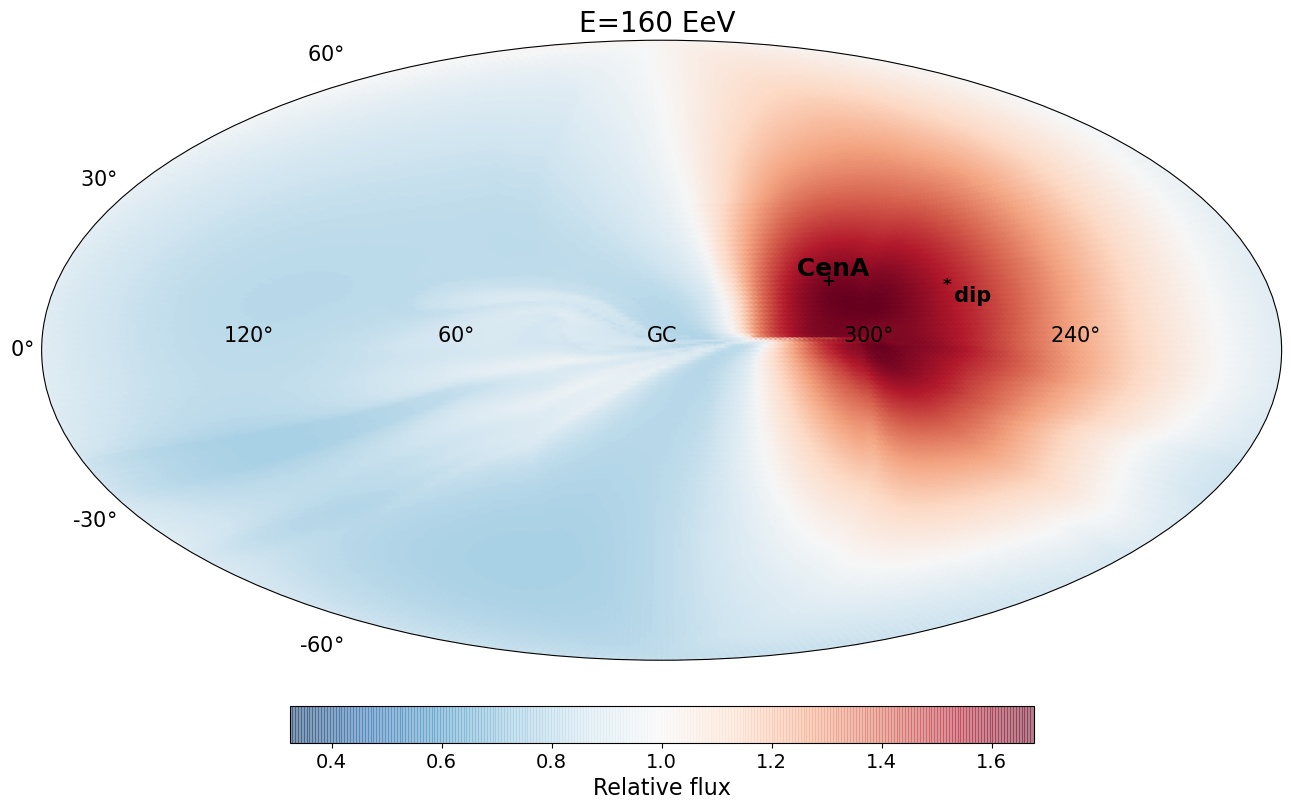}\\
  
    \caption{CR flux, normalised to the average flux in the whole sky,  as a function of galactic coordinates for the model G-NE-4Mpc(CO) with $\Delta=2$, $ct_i=50r_{\rm s}$ and $l_{\rm coh}=100$~kpc (considering the observed energies scaled by 1.14). Different panels correspond to different energies, as indicated, and the label `dip' indicates the direction of the resulting dipole. The location of Cen~A is also indicated.}
    \label{fGCOBmaps}
\end{figure}

To get an idea about the resulting anisotropies in these kind of scenarios, we show in Fig.~\ref{fGCOBmaps} the predicted distribution of arrival directions for the scenarios obtained with the positive systematic shifts in the energies which led to the smallest $\chi^2$. We show the maps for energies of 5, 10, 20, 40, 80 and 160\,EeV, for which the overall dipole amplitudes obtained are 2.7, 9.7, 17, 28, 47 and 32\% respectively. The dipoles point towards the directions indicated as `dip' in the plots. Also a more localised region of excess is observed near Cen~A. Due to the coherent Galactic field deflections, which we evaluate with the JF12 model  \cite{ja12}, this excess is slightly shifted towards the direction of the outer spiral arm, located at galactic coordinates ($\ell, b)\simeq (260^\circ, 0^\circ)$. The resulting flux distributions are in qualitative agreement with the observed ones, but a precise quantitative agreement would require to consider many more specific details, such as the nonomogeneity of the EGMF, the precise model of the Galactic  magnetic field, the detailed emission history of the source, the systematic effects on the energy or $X_{\rm max}$ measurements,  the actual shape of the cutoff suppression,  etc.

 \section{The He anisotropy}
 
As was clear from the results of previous sections, the He component should be strongly suppressed above 20\,EeV due to the source cutoff as well as by the interactions with the CMB.
In principle between 10 and 20\,EeV the He nuclei should have a similar intrinsic anisotropy as the CNO elements above 30 to 40\,EeV (depending on the actual element dominating this group). This is the regime in which an excess towards the Cen~A direction is observed at intermediate angular scales, with the maximum significance being found for top hat windows of 27$^\circ$ radius and energies above 38~EeV \cite{aa15,go23b}. Given the much larger CR fluxes present between 10 and 20\,EeV with respect to those above 40\,EeV, one could expect that the anisotropy of the He component could be  significant, unless its fractional contribution to the flux is small. We will now discuss what type of signatures could be expected in this region of the sky  associated to the He component  from the source (these type of considerations  are analogous to those discussed in \cite{le09,ab11} to constrain the  anisotropies from the H component in different source scenarios). 

 In the scenarios being considered here, the largest overdensity around the Cen~A direction that was observed above $E_{\rm th}=38$~EeV should result from the CNO component, which largely dominates the flux above this threshold (typically amounting to more than 80\% of the flux). Also, note that the heavier nuclei contributing the rest have on average a smaller  rigidity than the CNO ones and hence have a wider average angular spread. In this window, $N_{\rm obs}=237$ events were observed for an expectation of $N_{\rm exp}=169$ \cite{go23b}, corresponding to a relative excess of $\lambda=N_{\rm obs}/N_{\rm exp}\simeq 1.4$, leading to a significance of about $\Sigma(E>E_{\rm th})\equiv (N_{\rm obs}-N_{\rm exp})/\sqrt{N_{\rm exp}}=\sqrt{N_{\rm exp}}(\lambda-1)\simeq 5.2\sigma$. This significance becomes  4.0$\sigma$ after penalisation for the scan over threshold energies and angles. One may consider that the He component from the nearby source above the rescaled threshold $E_{\rm th}^{\rm He}\equiv (2/Z_{\rm CNO})E_{\rm th}$, with $Z_{\rm CNO}=6$ to 8 depending on the actual composition of the CNO group, has a similar relative excess as that observed above $E_{\rm th}$ where the CNO component from the nearby source dominates. Hence, in the same window one should have for the He contribution from the nearby source $N_{\rm obs}^{\rm s}({\rm He})\simeq \lambda N_{\rm exp}^{\rm s}({\rm He})$, and we can consider that above this lower energy threshold the overall anisotropy in that window of the heavier elements, as well as that of the LE component, are much smaller. In this case,  the significance associated to the excess of the He component should be
\begin{equation}
    \Sigma(E>E_{\rm th}^{\rm He})\simeq\sqrt{N_{\rm exp}^{\rm tot}(E>E_{\rm th}^{\rm He})}X_{\rm He}^{\rm s}(\lambda-1)\simeq \sqrt{\frac{N_{\rm exp}^{\rm tot}(E>E_{\rm th}^{\rm He})}{N_{\rm exp}^{\rm tot}(E>E_{\rm th})}}X_{\rm He}^{\rm s}\Sigma(E>E_{\rm th}),
\end{equation}
where $X_{\rm He}^{\rm s}$ is the fraction of the observed events above the threshold $E_{\rm th}^{\rm He}$ which are due to the He component from the nearby source (with  part of these He being secondaries from C/O disintegrations). Considering for instance the model with the CO emission at the source, for which $E_{\rm th}^{\rm He}\simeq 12.5$\,EeV, and using that from the ratio of the total fitted fluxes above the respective thresholds one has that $N_{\rm exp}^{\rm tot}(E>E_{\rm th}^{\rm He})\simeq 12 N_{\rm exp}^{\rm tot}(E>E_{\rm th}) $, one finds that one should expect 
\begin{equation}
   \Sigma(E>E_{\rm th}^{\rm He}) \simeq  \frac{X_{\rm He}^{\rm s}}{0.3}\Sigma(E>E_{\rm th})\simeq 2.6 \frac{X_{\rm He}^{\rm s}}{0.15}.
\end{equation}
 One should also keep in mind that this significance should not be subject to a penalisation, since the threshold and angular scale were already selected in the previous search. For the models with small values of $R^{\rm s}_{\rm cut}$, one finds a more prominent He source contribution, with  $X_{\rm He}^{\rm s}\simeq 0.2$, so that a marginally significant excess should be expected from the He source component. 
 On the other hand, the reduced source He contribution obtained in the scenarios with a larger values of $R^{\rm s}_{\rm cut}$, which typically lead to $X_{\rm He}^{\rm s}\simeq 0.1$, should lead to a smaller anisotropy associated to the He component from the source. Dedicated studies of the associated excesses at these lower energies can then significantly test the He contribution from the source in these scenarios. 
   Note also that the He source contribution  could  be reduced for instance if the $X_{\rm max}$ measurements were to have been affected by a systematic positive shift towards deeper showers, so that actually the true $\langle {\rm ln}A\rangle$ value should be larger than what considered here, or a similar change could also arise from some departures of the hadronic model from the one considered. 
 
 Finally, if a non-negligible H contribution were emitted from the source, it may also give rise to a localised anisotropy around Cen~A at an energy larger than about 40~EeV$/7\simeq 6$~EeV. On the other hand, the secondary H produced in photodisintegrations have a smaller rigidity than their parent nuclei (by a factor two), but in the scenarios with large source cutoff rigidities they may provide a contribution to the anisotropies at energies around $R^{\rm s}_{\rm cut}/2$.

\section{Conclusions}

We have considered the predictions for the spectrum and mass composition in a scenario in  which the nearby source Centaurus~A contributes most of the CRs above the ankle energy. Assuming a power-law injection with rigidity dependent cutoff, the maximum  energies of the CRs reaching Earth from this source turn out to be typically smaller than $10Z$\,EeV. For these moderate rigidities, the effects of the EGMF can produce a significant blurring of the source while the Galactic magnetic field can lead to lensing effects and multiple imaging which could make the arrival distribution of the UHECRs more in line with the observations \cite{mo19, mo22}. In order to get a level of anisotropies comparable to the one observed, the EGMF amplitude within the local neighbourhood containing the nearby source should be of few tens of nG and note that fields of this size are within those expected from equipartition of the magnetic energy with thermal gas motions within the filaments and sheets of the large scale structure \cite{ry98}

The obtained spectral index of the source, $\gamma_{\rm s}\simeq 1$ to 2 (depending on the assumed spectral steepening parameter $\Delta$ and possible systematic energy shifts), is closer to the expectations from diffusive shock acceleration than in the usual case in which a continuous distribution of sources is adopted and magnetic fields are ignored. The spectrum at Earth from a single source turns out to be  more sensitive to the particular elements injected than when many sources get averaged, and we found in particular that considering that the CNO group consists of a mixture of C and O nuclei provides a better description of the data than in the case in which just N nuclei are assumed to represent this group of elements. The inclusion of a Galactic component helps to explain the increase in the mass dispersion observed below a few EeV. 

The main suppression effect restricting the maximum energies of the different elements is often that due to the source rigidity cutoff, although in some cases, in particular when  systematic upward shifts on the measured energies are considered, the interactions during propagation may play an important role  even if the distance to the source is so small. In addition, the spectrum gets steepened by the transition from the diffusive to the quasi-rectilinear regimes and at low rigidities it becomes  suppressed due to the effect of the finite source emission time. 
The Si group elements can make a sizeable contribution close to the highest observed energies, with their spectrum being shaped by the interactions, and Fe group elements also contribute at the highest energies.
The relative Si/Fe source abundances is not strongly constrained with present data and its value should  affect the shape of the spectrum at the highest energies.  If no significant amounts of even heavier nuclei are emitted by the source, the future study of the shape at the end of the spectrum may help also to directly determine the source distance and to better probe this scenario. 

The Si and Fe nuclei above the suppression energy and close to the highest observed energies should show an anisotropy similar to that of the CNO nuclei above 40\,EeV, without the more isotropic small background that these heavy nuclei produce below the suppression energy. Anyhow, due to the very steep falloff of the spectrum at these energies it may still prove more fruitful to study the anisotropies in the region of the CNO suppression (40 to 80\,EeV) rather than 
above 100\,EeV, where the number of events is much scarcer. Also note that due to the heavy composition at the highest energies and the strong EGMF considered between us and the nearby source, the deflections of the highest energy events observed by different observatories remain large, explaining the apparent lack of observed close associations with nearby source candidates.
As we also discussed, if the excess observed above 38\,EeV around Cen~A is attributed to the CNO component from the nearby source, a signal due to the He component of that source should be expected at the correspondingly rescaled threshold of about 10 to 15\,EeV,  and  constraints on the significance of this excess may be used to restrict  the  fraction of He contributed by the nearby source, helping to select the best models. 

Another signal that could eventually help to identify the source is that of very high energy neutrinos produced in the source environment, whose spectrum may extend up to an energy $R_{\rm cut}^{\rm s}/40\simeq 0.2$ to 0.5\,EeV (i.e. the maximum nucleon energy divided by 20), or twice as large if a subdominant H component is present at the source. These could result from hadronic interactions with gas or by photopion production off optical/UV photons present around the source. On the other hand, no significant amounts of cosmogenic neutrinos will get produced by the CRs from this source in the interactions with radiation backgrounds as they propagate through intergalactic space, given the strong suppression of the photopion production off CMB or EBL photons at these rigidities and  the short distance involved. A comprehensive study of these scenarios should also allow one to learn about the EGMF strength in our neighbourhood  and about the emission history of the nearby source.

\section*{Acknowledgments}
We are grateful to Juan Gonz\'alez for help with the propagation matrices used. This work was supported by CONICET (PIP 2021-0565) and ANPCyT (PICT 2021-I-A-00547).

\end{document}